\documentclass[12pt,preprint]{aastex}
\usepackage{graphicx}
\usepackage{amsmath}
\usepackage{txfonts}
\usepackage{natbib}

\shorttitle{An improved catalog of halo wide binary candidates}
\shortauthors{Allen \& Monroy-Rodr\'{\i}guez}
\begin{document}

\title{An improved catalog of halo wide binary candidates}
\author{Christine Allen \altaffilmark{1}  \& Miguel A. Monroy-Rodr\'{\i}guez \altaffilmark{1}}
\affil{Instituto de Astronom\'{\i}a, Universidad Nacional Aut\'onoma de M\'exico, Apdo. Postal 70-264, M\'exico, D.F. 04510, M\'exico; chris@astro.unam.mx}

\begin{abstract}
We present an improved catalog of halo wide binaries, compiled from an extensive  literature search.  Most of our binaries stem from the common proper motion binary catalogs by Allen et al. (2004), and Chanam\'e \& Gould. (2004) but we have also included binaries from the lists of Ryan (1992) and Zapatero-Osorio \& Martin (2004). All binaries were carefully checked and their distances and systemic radial velocities are included, when available.  Probable membership to the halo population was tested by means of reduced proper motion diagrams for 251 candidate halo binaries.  After eliminating obvious disk binaries we ended  up with 211 probable halo binaries, for 150 of which radial velocities are available.  We compute galactic orbits for these 150 binaries and calculate the time they spend within the galactic disk.  Considering the full sample of 251 candidate halo binaries as well as several subsamples, we find that the distribution of angular separations (or expected major semiaxes) follows a power law $f(a) \sim a^{-1}$ (Oepik's relation) up to different limits.  For the 50 most disk-like binaries, those that spend their entire lives within  $z = \pm 500$~pc, this limit is found to be 19,000 AU (0.09 pc), while for the 50 most halo-like binaries, those that spend on average only 18\% of their lives within  $z = \pm 500$~pc, the limit is 63,000 AU (0.31 pc).  In a companion paper we employ this catalog to establish limits on the masses of the halo massive perturbers (MACHOs).
\end{abstract}

\section{Introduction}
Old disk and halo binaries are relevant to the understanding of processes of star formation and early dynamical evolution.  In particular, the orbital properties of the wide binaries remain unchanged after their formation, except for the effects of the galactic tidal field or of their interaction with perturbing masses encountered during their lifetimes, as they travel in the galactic environment. The widest binaries are quite fragile and easily disrupted by encounters with various perturbers, be they passing stars, molecular clouds, spiral arms, MACHOs (massive compact halo objects) or the galactic tidal field.  In this way, they can be used as probes to establish the properties of such perturbers (Weinberg, Shapiro \& Wasserman 1987; Wasserman \& Weinberg 1991;  Yoo, Chaname \& Gould 2004; Jiang \& Tremaine 2010)).

An interesting application of halo wide binaries was proposed by Yoo et al. (2004) who used the catalog of Chanam\'e \& Gould (2004, hereinafter ChG) containing 116 halo binaries to try to detect the signature of disruptive effects of MACHOs in their widest binaries.  They were able to constrain the masses of such perturbers to  $m < 43~ M_\sun$, essentially excluding MACHOs from the galactic halo, since a lower limit of about $30~M_\sun$ is found by other studies (Alfonso et al. 2003; Alcock et al 1998, 2001).  However, as was  shown a few years later by Quinn et al (2009), this result depends critically on the widest binaries of their sample, as well as on the observed distribution of the angular separations which, according to ChG, shows no discernible cutoff between $3.5\arcsec$ and $900\arcsec$.  Quinn et al (2009) obtained radial velocities for both components of four of the widest binaries in the ChG catalog and found concordant values for three of them, thus establishing their physical nature.  The fourth binary turned out to be optical, resulting from the random association of two unrelated stars.  Excluding this spurious pair and assuming power law distributions with exponents between $0.8$ and $1.6$ for the observed separations they find a limit  for the masses of the MACHOs of $m < 500~M_\sun$, much less stringent than that found previously;  they conclude rather pessimistically that the currently available wide binary sample is too small to place meaningful constraints on the MACHO masses.  They also point out that the density of dark matter encountered by the binaries along their galactic orbits is variable, and that this variation has to be taken into account when calculating constraints on MACHO masses.

Motivated by these concerns, we have constructed a catalog of 251 candidate halo wide binaries.  We describe the way the catalog was compiled in Section 2.  In Section 3 we present some statistical properties for different subsamples of binaries, and examine the galactic orbits for the 150 binaries with sufficient observational data.  Section 4 is devoted to a discussion of our results.  We point out that for a total of 9 binaries concordant radial velocities for both components were found in the literature.  A companion paper utilizes this catalog to dynamically model the evolution of wide halo binaries, and to obtain new limits on the MACHO masses.

We stress that our study pertains to the wide binaries only.  The classical study of Duquennoy \& Major (1991) and its recent update (Raghavan et al. 2010) shows a log-normal distribution of periods.  Our binaries correspond to the periods greater than the maximum period of these distributions, i.e. $P > 10^5$~yr, which corresponds approximately to semiaxes greater than 100~AU.  For our present purpose, we use the term ``wide binary'' to refer to binaries with separations larger than that corresponding to the maximum in the Kuiper or Duquennoy-Mayor distribution, which are presumably formed by processes different from those responsible for the closer binaries.

\section{Construction of the catalog}
A search of the literature was conducted, looking for high velocity, metal-poor wide binaries, since such samples are likely to be rich in halo stars.  Most of the binaries in the new catalog stem from the lists of Ryan (1992), Allen et al. (2000), Chanam\'e \& Gould (2004), and Zapatero-Osorio \& Martin (2004).  All listed data were checked, and updated when necessary.  We selected the most reliable data for the distances, metallicities and radial velocities.  All proper motions were checked in the Simbad database and non-common proper motion companions were eliminated (i.e., we omitted pairs with proper motions differing by more than the published values).  The catalog includes 111 halo binaries from Allen et al (2000), 110 halo binaries from Chanam\'e \& Gould (2004) 23 from Zapatero-Osorio \& Martin (2004) and 7 from Ryan (1992), to give a total of 251 halo binary candidates.

In order to refine the sample, we constructed a reduced proper motion diagram (RPM), following  Salim \& Gould (2003) and ChG (2004).  The RPM diagram for 251 candidate halo binaries is shown in Figure 1.   The black diagonal line is the limit  taken by ChG to separate disk from halo binaries.  We note that this line excludes quite a few binaries for which galactic orbits from Allen et al. (2000) clearly imply  their halo membership.  Therefore, we draw a new limit, indicated by the dotted line, which includes most of the Allen et al. binaries with halo-like galactic orbits.  Regardless of their position in the RPM diagram we exclude 40 binaries with galactic orbits such that they spend their whole lives within the disk, although their motions are clearly indicative of a thick disk or of an even dynamically hotter population.   In this way, we end up with 212 binaries, most likely to belong to the galactic halo or thick disk.

	To carry out the dynamical modelling of the evolution of halo binaries, the most useful quantity is the fraction of the lifetime that each binary spends far from the galactic disk.  Thus, galactic orbits were calculated for all binaries with available radial velocities (at least for their primaries), a total of 150 systems.  The Allen \& Santill\'an (1991) galactic potential was used, and the orbits were calculated backwards in time for 13 Gyr.  This allowed us to compute for each system $t_d/t$, the fraction of time spent within $z = \pm 500$~pc.

	The entire catalog is presented in Table 1. The entries are ordered by right ascension (J2000).  Distances to the primaries were taken, when appropriate, from the  Hipparcos catalog.  Hipparcos distances were adopted for those stars with relative errors in their trigonometric parallaxes of less than 15\%.  For stars with larger errors, photometric distances were preferred, mostly taken from the lists of  Nissen \& Schuster (1991) and and references therein, or by means of the polynomial used by ChG, adopting their photometry.    For a few stars not found in those lists, spectroscopically derived distances were taken from either Ryan (1992) or Zapatero-Osorio \& Martin 2004). Thus, most of the binaries listed in the catalog should have errors of less than 15\% in their adopted distances.

	Absolute visual magnitudes for the primaries were calculated from the adopted distances and the apparent visual magnitudes given in the respective catalogs.  Absolute magnitudes for the secondaries were calculated as described in Allen et al. (2000) for their stars.  They are mostly based on accurate photometry for the primary, and magnitude differences in two colors for the secondaries, a procedure which was shown by Allen et al. (2000) to yield reliable values for the absolute magnitudes of the secondaries.  For binaries stemming from Chanam\'e \& Gould their photometry was adopted. Secondaries from other sources are listed as in the original references.

	Linear separations between the components were calculated from the  angular separations and the adopted distances.  The linear separations were transformed into expected values for the major semiaxes of the binaries by the statistical relation (Couteau 1960)

\begin{equation}
 \langle a\arcsec \rangle = 1.40 s\arcsec . \nonumber\\
\end{equation}

	Therefore, a binary with projected angular separation $s\arcsec$ will have an expected major semiaxis $\langle a\rangle$  given by

\begin{equation}
\langle a \rangle ({\rm AU}) = \frac{1.40}{\pi} s\arcsec . \nonumber
\end{equation}

The above formula was derived by Couteau on theoretical grounds, assuming a random distribution of the geometrical elements of the orbit, as well as random times of passage through periastron. This relation turns out to be independent of the eccentricity distribution. Couteau also checked its observational validity based on more than 200 orbits, for which the constant turned out to be 1.411. A more recent empirical study (Bartkevi\v{c}ius 2008), based on data from the Catalog of Orbits of Visual Binary Stars (Hartkpof \& Mason 2007) and the WDS (Mason et al. 2007). The result was $\log~\langle a \rangle - \log s$ ranging from $0.112$ to $0.102$. In view of the large uncertainties of the empirically derived values, and to facilitate comparisons with previous work, we shall use Couteau's theoretical result.

	In Table 1, we list in Column 1 the NLTT number of the primary, as well as the designation of its companion.  In Column 2 we list alternative identifications of the primary.  We provide, in order of preference, identifications in the Hipparcos, the HD, the Giclas (G) or the Wilson catalogs.  Column 3 contains the identification of the secondary. Column 4 lists our best choice for the distance to the primary, as explained above.  Columns 5 and 6 list the absolute magnitudes of the primary and secondary, respectively. Columns 7 and 8 the projected angular separation between the components and the expected value for the major semiaxis.  The peculiar velocity of the binary is given in Column 9.  This velocity is computed assuming a solar motion of (9, 12, 7) km s$^{-1}$.  The radial velocity used to compute the peculiar velocity in Column 9 is taken from various sources, listed mostly in the Simbad database.
The presence of undetected close companions to one or both members of the binary will affect the radial velocities, and thus the computed orbits.  The observers have, in general, taken care to detect variations due to unseen companions. But the stars are faint and metal-poor, and thus observationally difficult.  For the orbit computations we checked the published radial velocities to make sure we were using the systemic radial velocity -when available.  Otherwise, we checked for radial velocity variations in the published data, and excluded a few systems with widely discrepant values.
Columns 10 to 12 contain the main orbital parameters of the galactic orbit;  we list the apocentric distance $R_{\rm max}$, the maximum distance from the galactic plane $|z_{\rm max}|$ and the three-dimensional eccentricity $e$ of the galactic orbit. In Column 13 we list $t_d/t$, the fraction of its lifetime the binary spends within the galactic disk ($z = \pm 500$~pc). In the last column, the provenance of the binary is given (A = Allen et al. 2000; CG = Chaname \& Gould, 2004; Z = Zapatero-Osorio et al., 2004; R = Ryan 1992).

\section{Some statistical properties}
\subsection{The distribution of absolute magnitudes}
Figure 2  shows the frequency distribution of the absolute magnitudes of the primaries (full line) and the secondaries (dotted line) of our catalog.  Note particularly the extension to the faint end of the magnitudes of the secondaries. This constitutes a sampling of the faint end of the main sequence for these metal-poor and presumably old stars. Observationally, these secondaries are challenging objects, being faint and having only a few lines in their spectra, but undoubtedly they are interesting and worthy of being included in observational programs.

\subsection{The distribution of peculiar velocities}
Figure 3 is a histogram of the distribution of the peculiar velocities of the binaries.  Note that most of the systems have peculiar velocities larger than 60 km s$^{-1}$, and thus can be considered to be good candidates to be bona fide thick disk or halo stars.

\subsection{The distribution of angular separations and expected major semiaxes}
Figure 4 shows the cumulative distribution of the angular separations of the full sample of 251  low-metallicity high-velocity binaries.  The straight line is a fit to the Oepik (1924) distribution  $f(s) = k/s$.  Figure 5 shows the same fit for the expected major semiaxes.  In Allen et al.(2000, 2012), as well as here, we choose to display the distributions in cumulative form, since in this way we can apply directly the Kolmogorov-Smirnov (KS) test to quantitatively assess the probability that the observed distribution stems from a given theoretical distribution.  Our choice of presentation of the angular distributions was criticized by ChG, who found a different exponent for the power law of their angular distributions.  Since the matter remains controversial, we present below a brief discussion of the issue.

In Figure 6 we show the cumulative distribution of the ChG sample of 110 halo binaries.  The straight line represents Oepik's distribution, corresponding to a power-law exponent of $-1$.  In Figure 7 we show the result of applying the Kolmogorov-Smirnov test to this sample.  It is obvious that Oepik's distribution shows an excellent fit to these stars up to an angular separation of 63 arc seconds, which includes 97 out of their 110 stars, leaving out those with the largest separations.  We have performed the same test using an exponent of $-1.6$, as ChG do, and find a good fit, but only for binaries with separations larger than 3 arc seconds (94 binaries). Thus, it would appear that this controversy cannot be settled merely by goodness-of fit arguments.  We would interpret the departure from Oepik's distribution as due to dynamical effects, an interpretation  supported both from theoretical studies (Valtonen 1997 showed Oepik's distribution to correspond to dynamical equilibrium) and from the observed distributions found for many independent samples of stars, and for groups of different ages (Poveda  et al. 2006.  Sesar et al. 2008; Lepine \& Bongiorno 2007: Allen et al. 2000).  We think the Oepik distribution is to be preferred for the following reasons.
\begin{enumerate}

\item It has been shown to hold for many different samples of wide binaries ($a \geq 100$~AU) stemming from different and independent sources (Poveda  et al. 2007,  Sesar et al. 2008; Lepine \& Bongiorno 2007; Allen et al. 2000).  We show below that both the angular separations and the expected semiaxes of the candidate halo binaries in our improved catalog  follow Oepik's distribution.  We also show that different subgroups of the catalog also follow Oepik's distribution up to different limiting semiaxes.

\item It has been shown to hold up to different limiting separations for wide binaries of different ages (Poveda \& Allen 2004;  Poveda et al. 2006) which is exactly what one would expect from dynamical considerations.

\item It holds for a volume-complete sample of nearby wide binaries (Poveda \& Allen 2004; Poveda et al. 2007) in the interval 133 AU $< a < 2640$ AU.

  \item It is thought to represent the primordial distribution of separations for wide binaries  both from theoretical reasons (Valtonen 1997) and from observations (it holds for the youngest binaries up to their largest separations, see Poveda \& Hern\'andez-Alc\'antara 2003; Poveda et al. 2006).

\item We have shown above that it holds for tha ChG halo binaries.  Indeed, we found it holds also for their disk binaries, and explains in a natural way, as the result of dissolution effects, the flattening of their distribution, which ChG found puzzling, and for which they offered no explanation.

\end{enumerate}

	  The cumulative frequency distribution of the angular separations for the more halo-like sample of 211 binaries is shown in Figure 8, that of the expected semiaxes in Figure 9. It is clearly seen that they follow Oepik's distribution. We performed KS tests to support this assertion, and found that Oepik's distribution holds up to angular separations of 160 arc seconds, and to expected semiaxes of 25~000~AU (0.12 pc). The latter test is shown in Figure 10.

	To further refine our sample of halo candidates we integrated galactic orbits for all binaries with available radial velocities (150), and calculated the times they spend in the galactic disk, that is, between $z$-distances of $\pm 500$~pc.  Figures 11, 12 and 13 display the meridional orbits of three representative binaries.  The cumulative frequency distribution of expected major semiaxes for this sample is shown in Figure 14, the result of the KS test in Figure 15.  The preceding  figures clearly show that Oepik's distribution holds for all these subsamples, although it does so up to different values of the separations or major semiaxes.

	Next, we separated the 150 binaries into three groups (each with an equal number of binaries), according to the time they spend within the galactic disk, as defined above.  We refer to these groups as the most disk-like, the intermediate, and the most halo-like binaries.  The average fraction of their lifetime the 50 most disk-like binaries spend in the disk is 100\%,  that of the most halo-like binaries is 18\%. To maximize the contrast we disregarded the intermediate group.

	Figures 16 and 17 show the cumulative frequency distribution of the expected major semiaxes for the most disk-like and the most halo-like groups.  The corresponding KS tests are displayed in Figures 18 and 19.  We see that the most disk-like binaries follow Oepik's distribution up to an expected major semiaxis of 19,000~AU (0.09 pc), while the most halo-like binaries do so up to an expected semiaxis of 63,000~AU (0.31 pc).

	These results confirm and reinforce the conclusions obtained in Poveda et al. (1997), Allen et al. (2000)  and Poveda \& Allen (2004), namely that (a) Oepik's distribution is followed by many different samples of wide binaries; (b) that the maximum semiaxis up to which Oepik's distribution holds is a function of the age of the group studied, as well as of the environment encountered by the binary along its galactic trajectory.  Thus, for example, Poveda et al. (1994) were able to divide the binaries in their catalog in two groups, ``probably young'' ($2 \times 10^9$~yr) and ``probably old'' ($> 4.6 \times 10^9$ yr), according to a variety of age indicators. Poveda \& Allen (2004) found Oepik's law to be valid up to semiaxes of 8,000~AU for the young group, and up to only 2400~AU for the old group. For a very young ($10^6$ yr) independent group of binaries in the Orion Nebula Cluster: Poveda \& Hern\'andez-Alcantara (2003) found Oepik's distribution to hold up to 45,000~AU.

The departure from Oepik's distribution can be interpreted as due to the effects of dynamical perturbations, which tend to decrease the binding energy and thus increase the separation of the binary, until its ultimate dissolution.  These effects, with the passage of time, will tend to eliminate the widest systems.  Weinberg et al. (1987), Wasserman \& Weinberg (1991) and more recently Jiang \& Tremaine (2010) have modeled the evolution of disk binaries subject to perturbations by passing stars and molecular clouds.  The former authors find galactic tides to have negligible effects for binaries with $a < 0.65$~pc, the latter take tides into consideration for their widest binaries ($a > 10^5$~AU). Our widest binaries would thus appear to be stable against galactic rides. The results of Weinberg et al. were found to be consistent with the distributions found for the youngest and oldest systems of the solar vicinity (Poveda et al. 1997).  For the samples studied here, the departure from Oepik's distribution sets in at much larger values than those obtained for even the oldest binaries of the solar vicinity.  Similarly as was done in Allen et al. (2000) we interpret these larger values as due mainly to the relatively small time spent by our binaries within the galactic disk, where most of the perturbations occur.  The  most halo-like binaries in Allen et al. (2000) spent on average 26\% of their lifetimes within the disk, and were found to follow Oepik's distribution up to $\langle a\rangle  = 20,000$~AU (0.1 pc).  The most halo-like binaries of our improved catalogue spend on average only 18\% of their lifetime within the disk and follow Oepik's distribution up to 63,000~AU (0.31 pc).

The question arises as to the effects of observational uncertainties  on our results.  Uncertainties in the distances will affect directly the inferred semiaxes, but not their distribution.  The galactic orbits will be affected by uncertainties in the distances, proper motions and radial velocities.  We have estimated the errors in the orbital parameters listed in Table 1 for a representative sample of binaries.  To this end, we computed for each binary two extreme orbits by adding and subtracting all the uncertainties to the ``central'' values.  This will provide an overestimate of the expected errors in the orbital parameters.  The average resulting errors for this sample turned out to be 18\% for $R_{{\rm max}}$, 14\% for $z_{{\rm max}}$ and 16\% for $e$. Binary membership into the  most halo-like and the most disk-like groups remained unchanged.

\section{Discussion and conclusions}

By means of an extensive literature search we have compiled a list of 251 candidate halo wide binaries and present it in Table 1. Proper motions, radial velocities, photometric data and distances were carefully checked for each system. We found that the distribution of separations and major semiaxes for all the subgroups studied follows Oepik's up to different limiting semiaxes.  We computed galactic orbits for 150 binaries and obtained the fraction of their lifetimes spent within the galactic disk.  Separating the binaries into three groups, most disk.like, intermediate and most halo-like we find that the most disk-like binaries begin to depart from Oepik's distribution at an $\langle a\rangle  = 19,000$~ AU, whereas the most halo-like do so at an $\langle a\rangle  = 63,000$~AU.

The great majority of the wide binaries in our catalog stem ultimately from Luyten's NLTT. The NLTT is quite complete up to visual magnitude 19, and for proper motions larger than 180 mas/yr. As such, it is a magnitude limited sample, and subject to the usual biases of such samples, namely, Malmquist, kinematical, and Lutz-Kelker biases. The rNLTT (Gould \& Salim 2003; Salim \& Gould 2003) covers the intersection of the First Palomar Observatory Sky Survey and the Second Incremental Release of the Two Micron Sky Survey, about 45\% of the sky.  As discussed by ChG, it is not necessary to have a volume-complete catalog, but a representative sample, with well understood selection effects.  Our catalog, as well as previous ones, is far from complete.  Certainly, many of the widest pairs will be missing. However, our main results on the distribution of semiaxes (showing the destructive effects of perturbers at large semiaxes), are not likely to be affected by incompleteness or biases.  Oepik's distribution was previously found to hold for a volume-complete sample of nearby wide binaries (Poveda \& Allen 2004, Poveda et al. 2007) in the interval  133 AU $< \langle a\rangle < 1640$~AU.  The same distribution was also found to hold (up to different values of the semiaxis) for different, independent samples of binaries, stemming from the LDS, the IDS amd the catalog of nearby wide binaries of Poveda et al. (1994).  Using a variety of age indicators, the latter binaries were separated into young and old groups;  the limiting $\langle a\rangle$ turned out to be 2400 AU for the oldest group and 7900 AU for the youngest group. Oepik's distribution was also found to hold up to $\langle a\rangle = 30~000$~AU  for a group of extremely young ($10^6$~yr) wide binaries in the Orion Nebula Cluster (Poveda \& Hern\'andez-Alc\'antara 2003). It was also shown to hold for the disk and halo binaries of ChG (Poveda et al. 2007).  We should mention that the Oepik distribution of separations was also found by Lepine \& Bongiorno (2007) and Sesar (2008) in their catalogs.  The consistency found between our present results and those previously found for many different and independent samples of binaries gives us confidence that incompleteness and selection effects are not significantly influencing our conclusions.

The origin of wide binaries has been a long-standing problem.  Several mechanisms have been proposed for this problem, among which we can mention the dynamical unfolding of triple systems (Reipurth \& Mikkola 2012), formation by capture of unbound stars in dissolving star clusters (Kouwenhoven et al. 2010, Moeckel \& Clarke 2011, Perets \& Kouwenhoven 2012), and the suggestion that they may be the sole remains of former moving groups or accreted structures in the galactic halo (Allen et al.2007). Some of these scenarios have as a result a thermal distribution of eccentricities, and an Oepik-like distribution of semiaxes, and would thus accord better with the empirical findings on the distribution of separations of binaries found in different environments and having different ages.  It is not clear which of these mechanisms -if any- would work for the halo binaries, but at this stage the scenarios involving cluster dissolution would seem more plausible, since they result in a thermal distribution of eccentricities and-in some cases- an Oepik-like distribution fo semiaxes

Our present results are consistent with those found in Allen et al. (2000) with a smaller sample. In a companion paper we show that they allow us to obtain better estimates of the masses of the halo perturbers, and they should be also useful for other dynamical studies.

Acknowledgments.
We thank A. Poveda, whose long-time interest in wide binaries inspired this work.  MAM is grateful to UNAM-CEP for a graduate fellowship. Our thanks are also due to M. Hern\'andez-Cruz for valuable assistance. This research used the Simbad database operated at CDS, Strasbourg, France.

\clearpage

\begin{figure}[tbh]
\includegraphics[width=15cm,height=15cm]{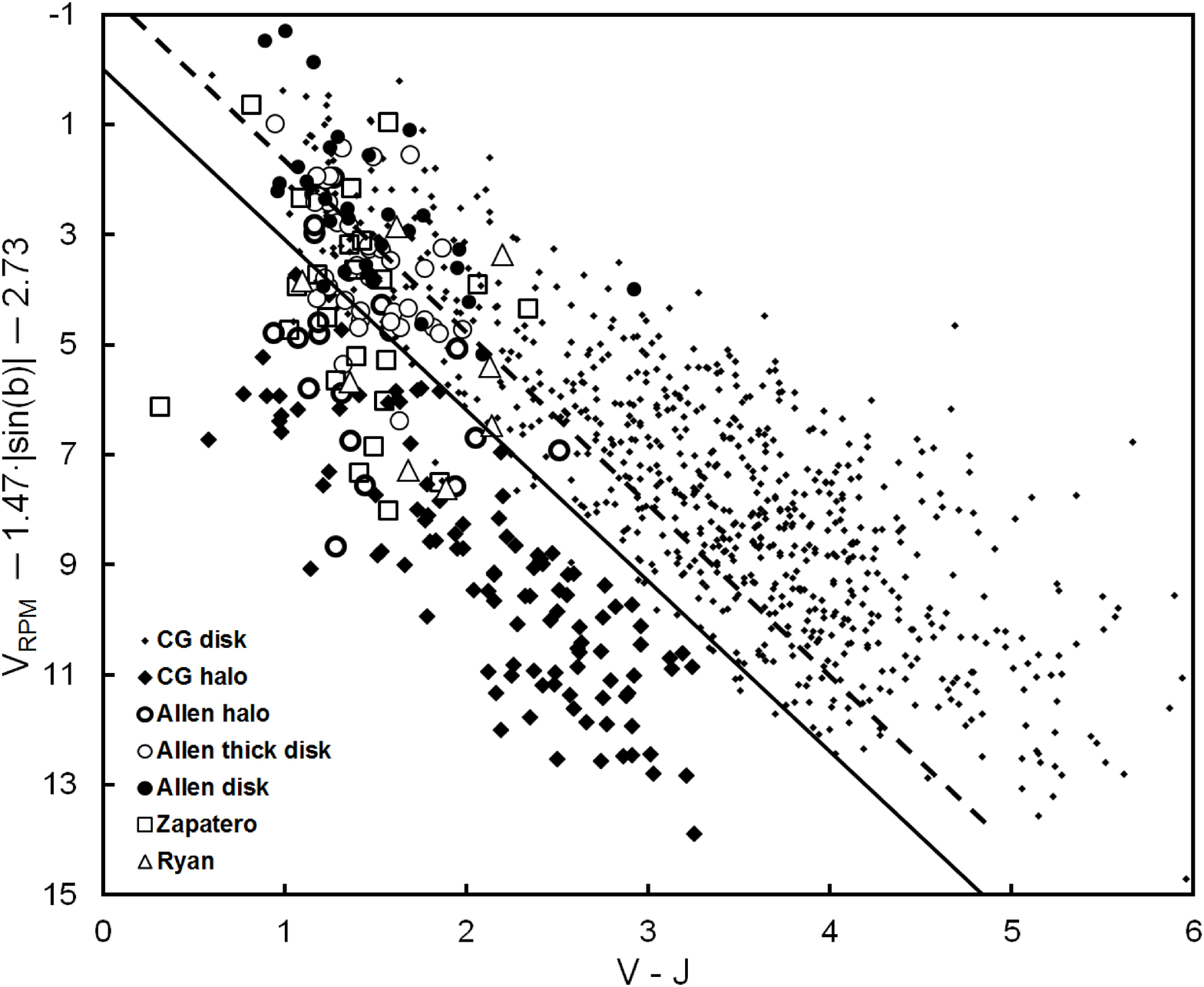}
\figcaption[]{Reduced proper motion diagram for 251  halo wide binary candidates. The axes are defined as in ChG (their eq. 2). The V-J plotted is that of the primaries. The vertical axis plots the reduced proper motion adjusted for galactic latitude $b$. Here $V_{{\rm RPM}} = V + 5 \log \mu$, with $V$ the $V-$magnitude and $\mu$ the proper motion. The symbols denote the provenance from different catalogues.  The full line was used by ChG to separate disk from halo binaries.  However, this line excludes quite a number of binaries whose galactic orbits clearly indicate membership to the halo as was shown in Allen et al. (2000). To identify candidate halo binaries we shifted the ChG line so as to include these stars, as shown by the dashed line.  Final halo membership was determined by computing the time the binaries spend within the galactic disk.  See text for details}
\end{figure}
\clearpage

\begin{figure}[tbh]
\centering
\includegraphics[width=8cm,height=8cm]{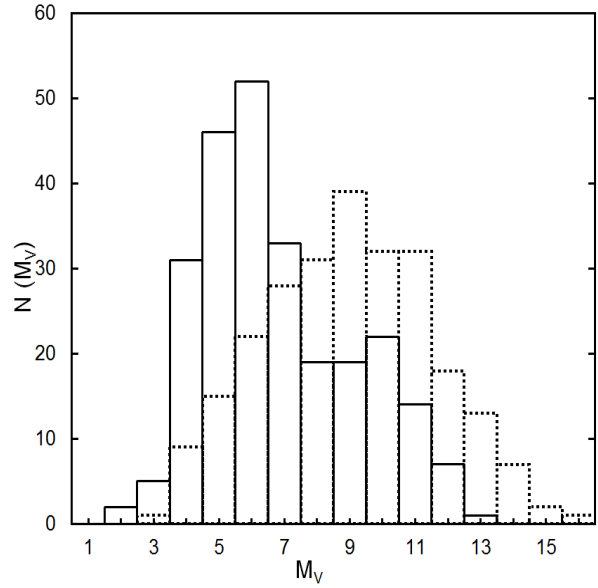}
\figcaption[]{Frequency distribution of the absolute magnitudes of 251 halo wide binary candidates.  The primaries (full line) and secondaries (dashed line) are shown.  Note the extension of the magnitudes of the secondaries to the faint end of the diagram.}
\end{figure}

\begin{figure}[tbh]
\centering
\includegraphics[width=8cm,height=8cm]{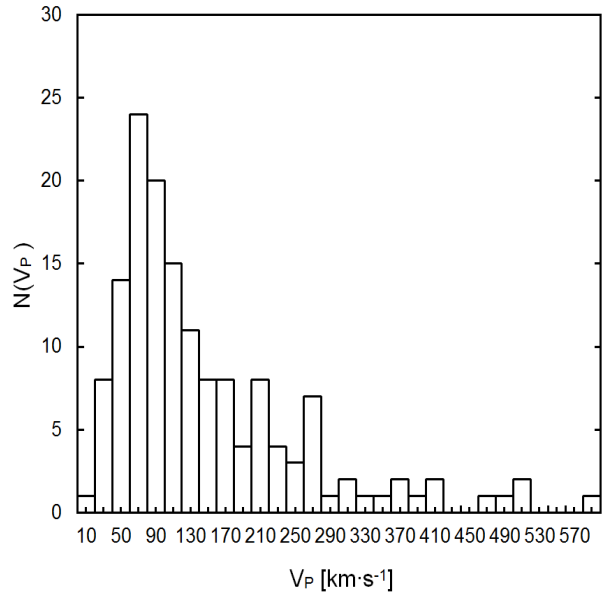}
\figcaption[]{Frequency distribution of the peculiar velocities of 150 binaries with available radial velocities.  Note that the majority of these stars have peculiar velocities much  larger than 65~km~ s$^{-1}$, clearly indicating that halo binaries predominate in this sample.}
\end{figure}

\begin{figure}[tbh]
\centering
\includegraphics[width=8cm,height=8cm]{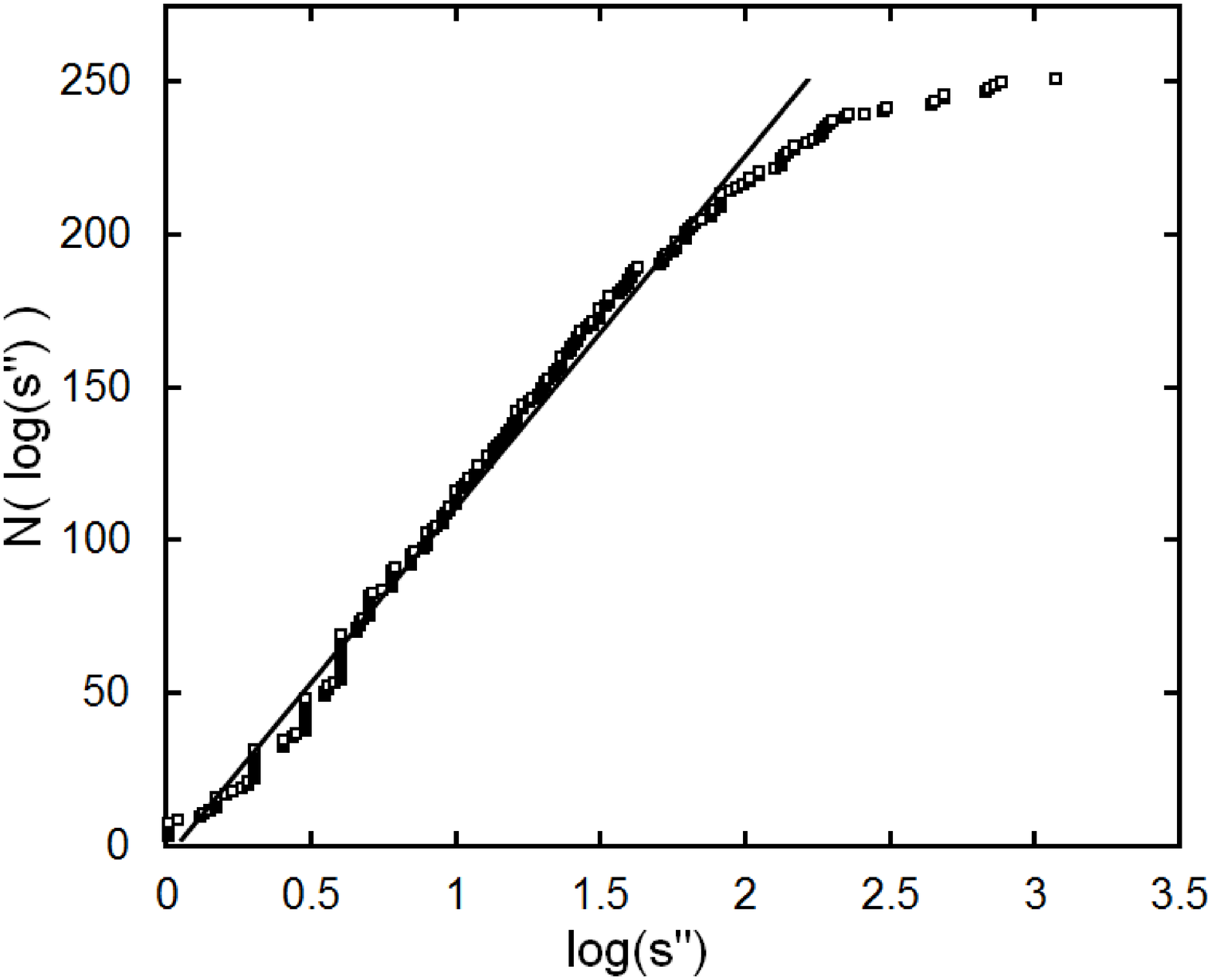}
\figcaption[]{Cumulative distribution of the angular separations of 251 halo wide binary candidates. The full line is a fit to the Oepik distribution $f(s) =$~k/s.}
\end{figure}

\begin{figure}[tbh]
\centering
\includegraphics[width=8cm,height=8cm]{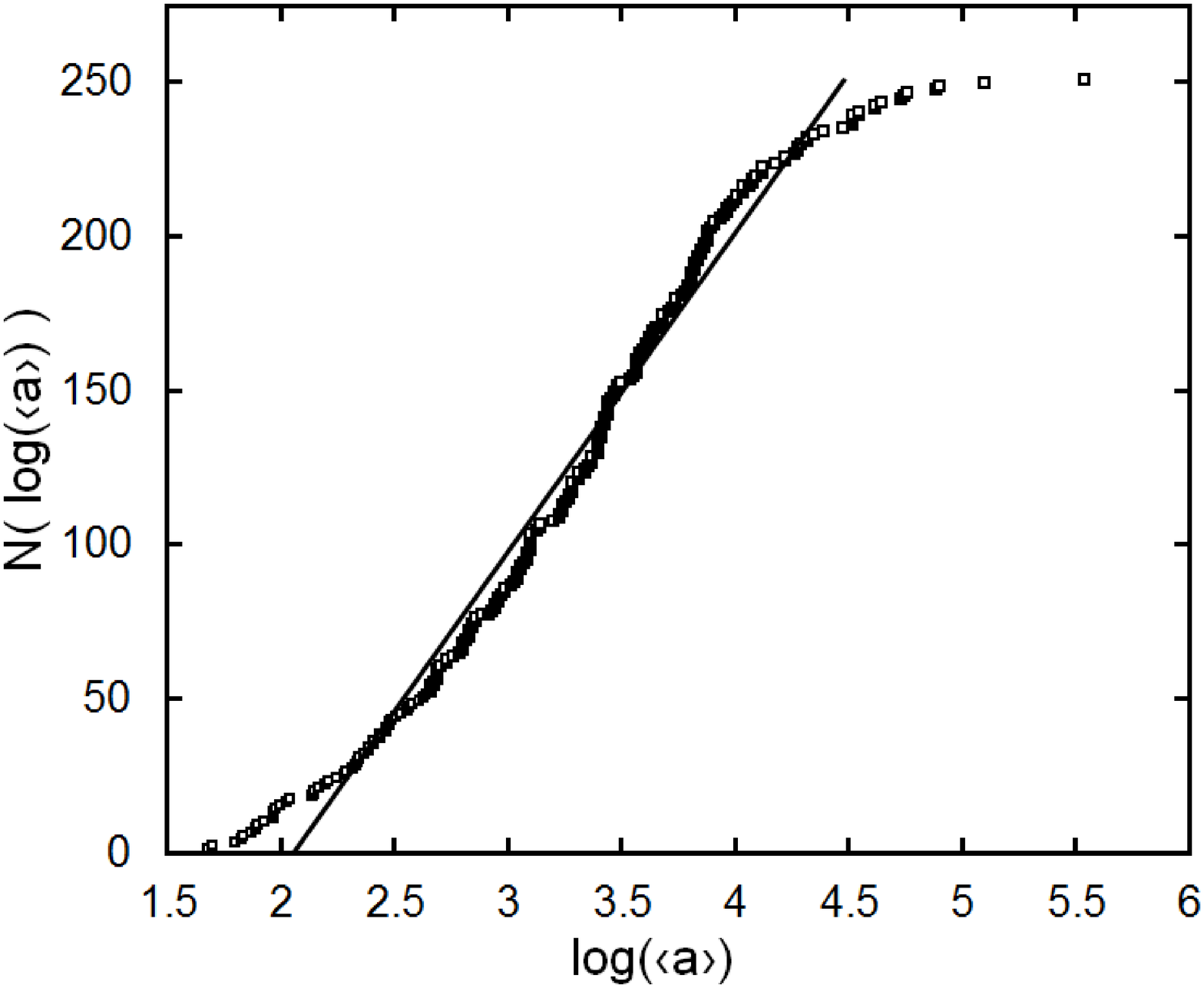}
\figcaption[]{Cumulative distribution of the expected major semiaxes of 251 halo wide binary candidates. The full line is a fit to the Oepik distribution $f(\langle a\rangle) =k \langle a\rangle^{-1}$. Units for $\langle a\rangle$ are AU.}
\end{figure}

\begin{figure}[tbh]
\centering
\includegraphics[width=8cm,height=8cm]{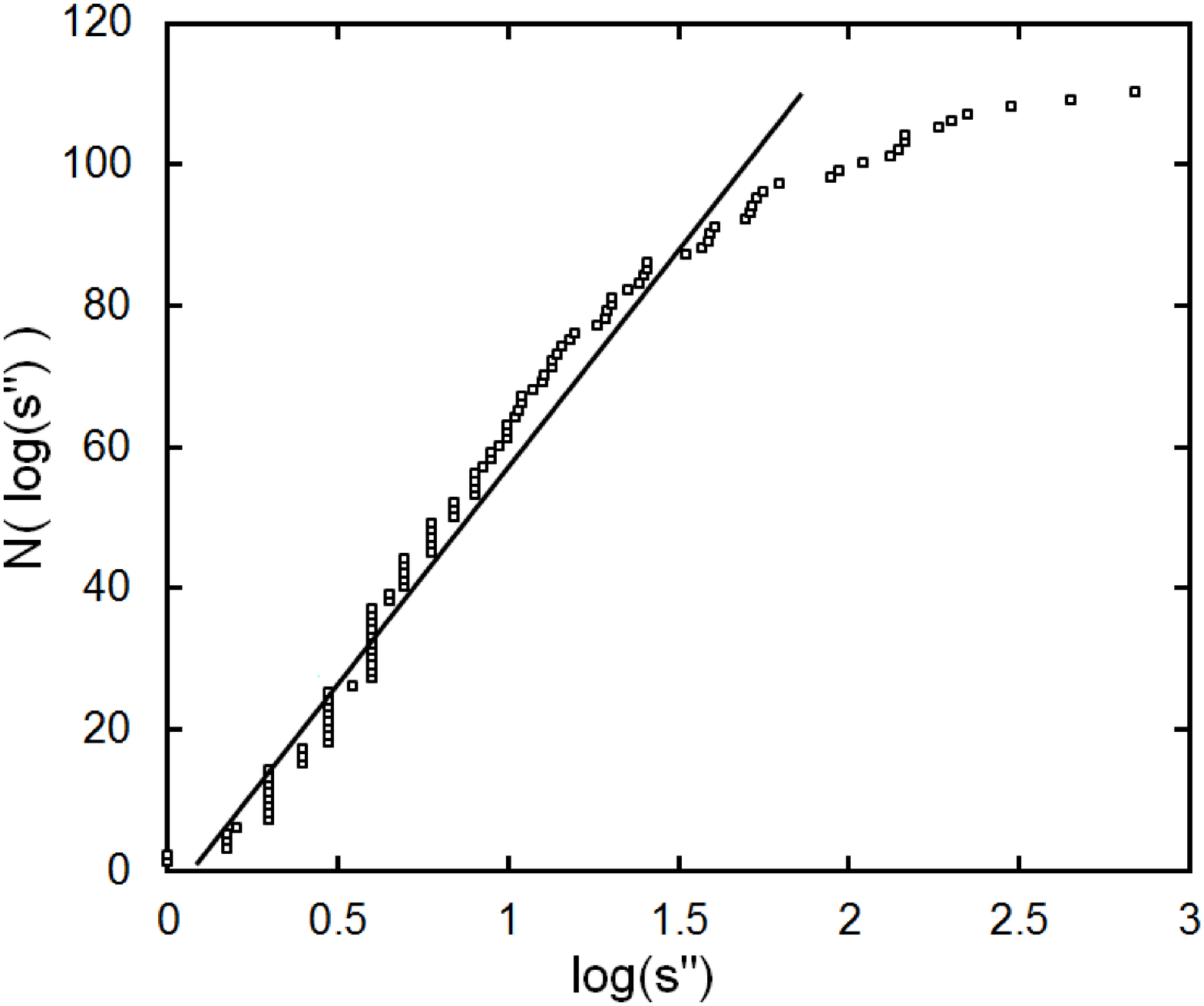}
\figcaption[]{Cumulative distribution of the angular separations of the 110 ChG  halo wide binaries.  The full line is a fit to the Oepik distribution $f(\langle a\rangle) =k \langle a\rangle^{-1}$.  The fit is seen to satisfactorily represent this sample.}
\end{figure}

\begin{figure}[tbh]
\centering
\includegraphics[width=8cm,height=8cm]{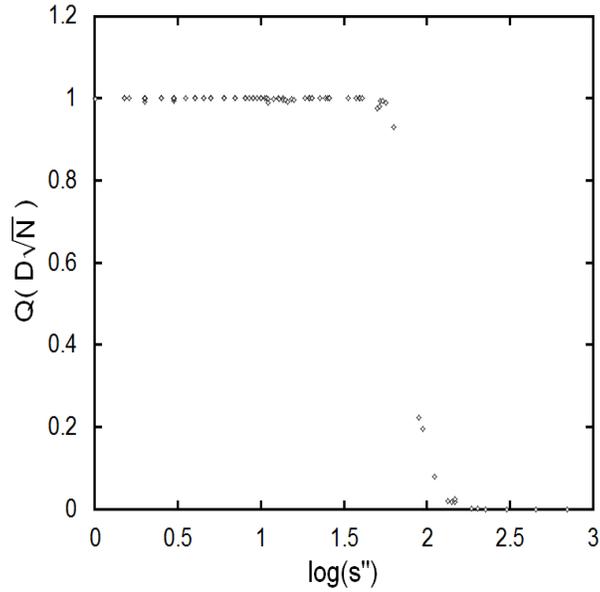}
\figcaption[]{Result of applying the Kolmogorov-Smirnov test to the binaries of ChG shown in Figure 6. The coefficient of significance $Q$ is plotted as the ordinate. The test shows that the fit is excellent  up to a separation of 63 arc seconds (97 binaries), where it abruptly deviates from the Oepik distribution.  See text for details.}
\end{figure}

\begin{figure}[tbh]
\centering
\includegraphics[width=8cm,height=8cm]{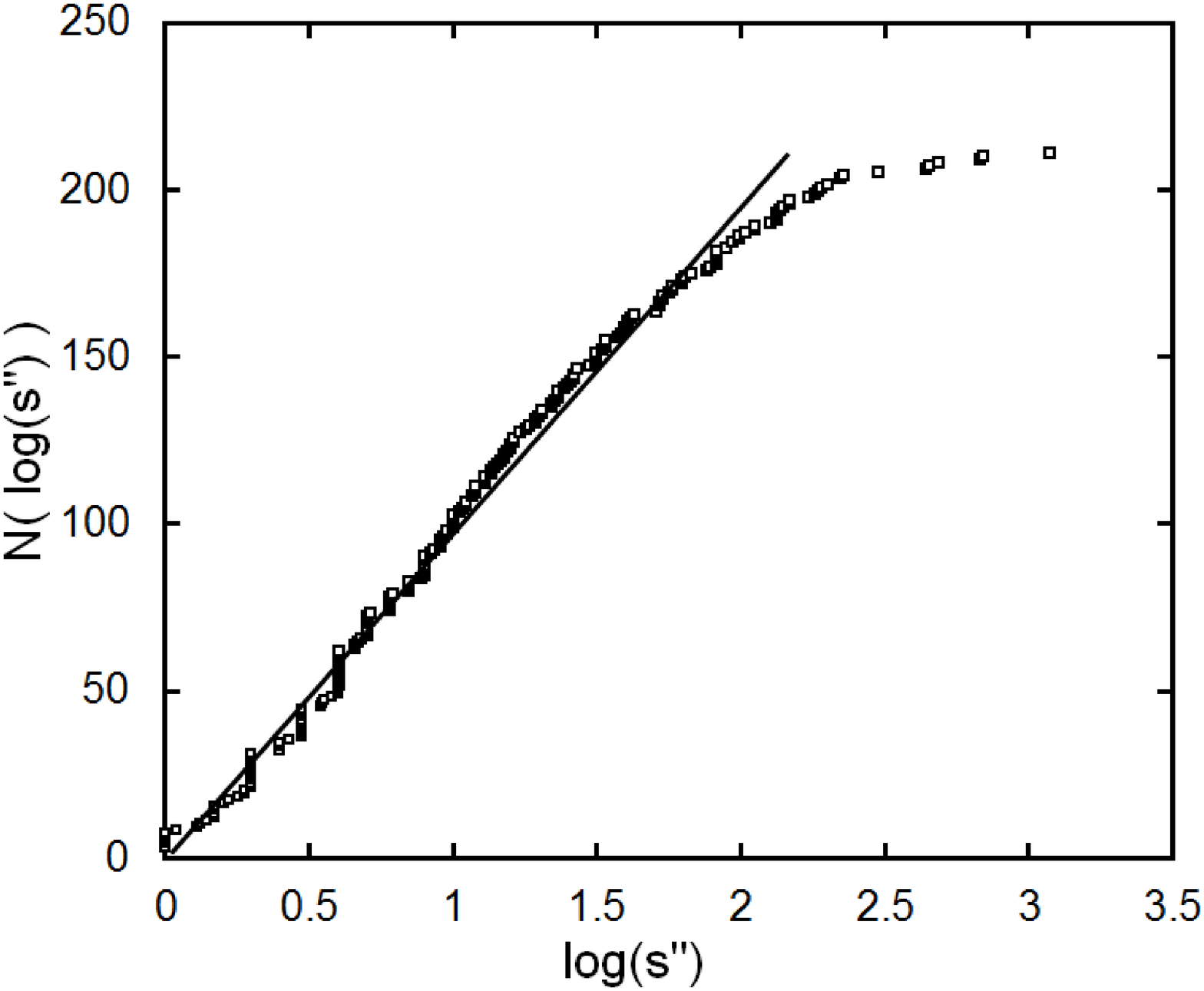}
\figcaption[]{Cumulative distribution of the angular separations of the sample of 211 halo wide binary candidates. The full line is a fit to the Oepik distribution $f(s) = k/s$.}
\end{figure}

\begin{figure}[tbh]
\centering
\includegraphics[width=8cm,height=8cm]{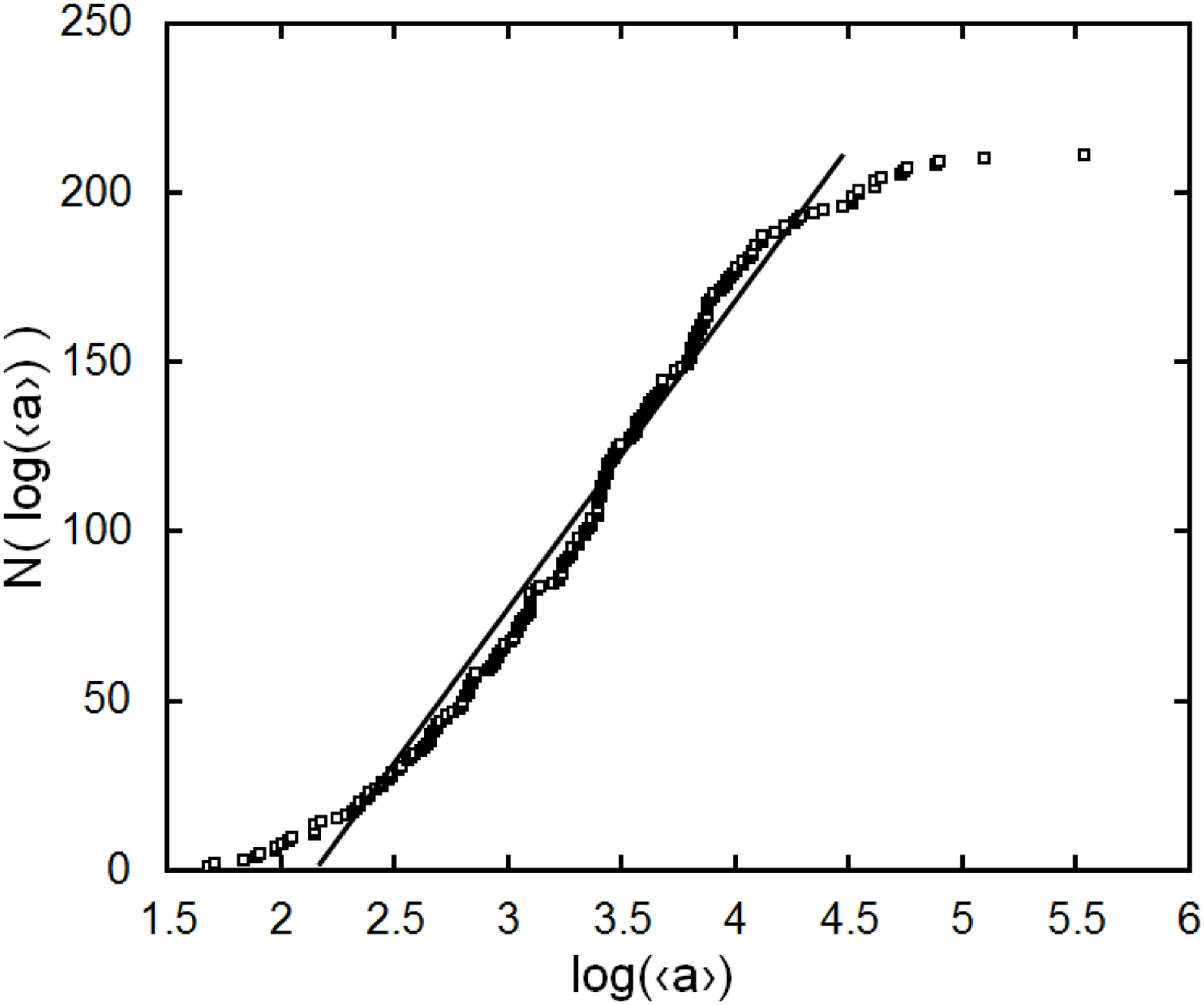}
\figcaption[]{Cumulative distribution of the expected major semiaxes of the sample of 211 halo wide binary candidates. The full line is a fit to the Oepik distribution $f(\langle a\rangle) =k \langle a\rangle^{-1}$. Units for $\langle a\rangle$ are AU.}
\end{figure}

\begin{figure}[tbh]
\centering
\includegraphics[width=8cm,height=8cm]{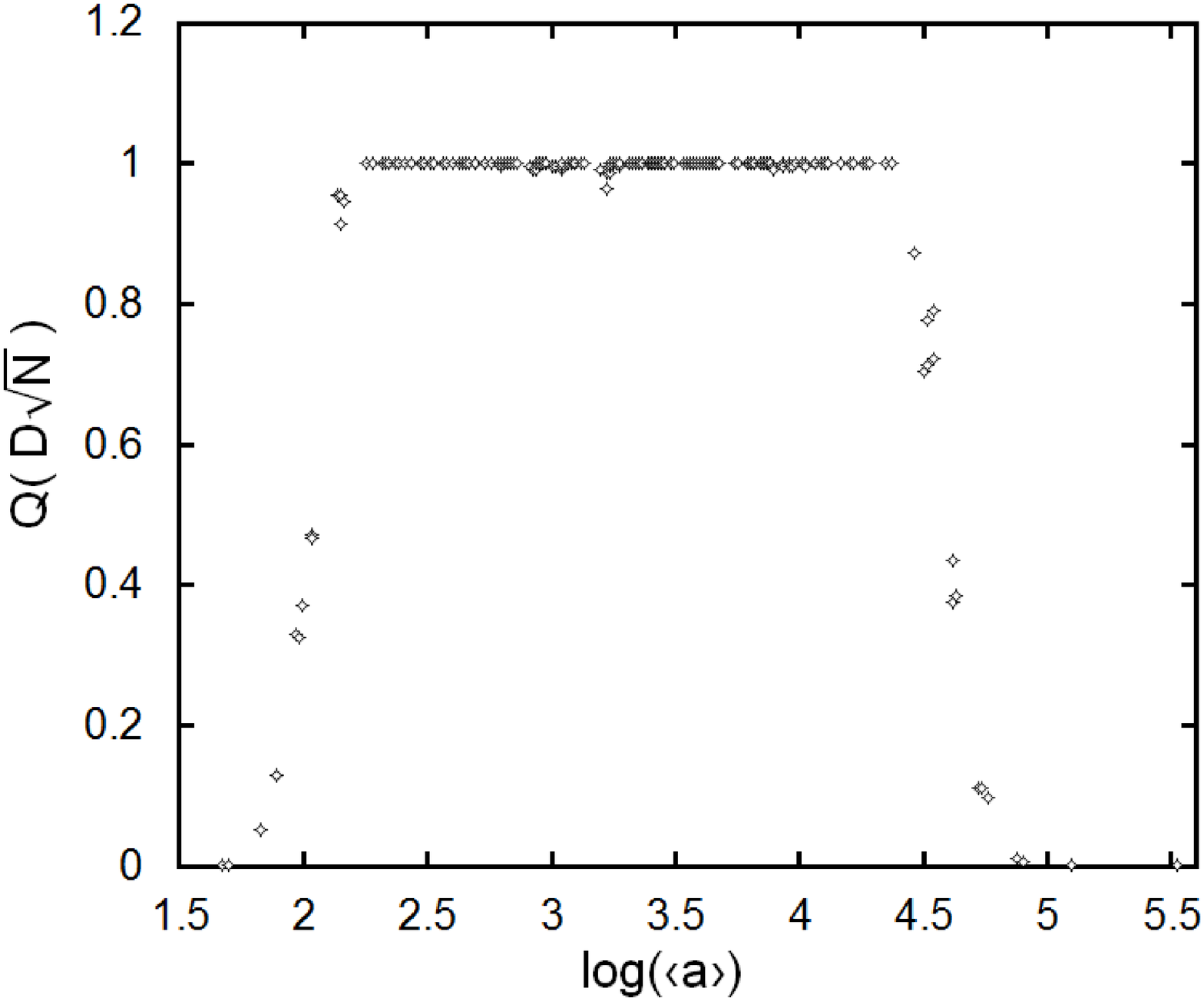}
\figcaption[]{Result of the Kolmogorov-Smirnov test for the 211 binaries of Figure 9.  Units for $\langle a\rangle$ are AU. The test shows that the Oepik ditribution holds up to an expected major semiaxis of 25 000 AU.}
\end{figure}

\begin{figure}[tbh]
\centering
\includegraphics[width=8cm,height=8cm]{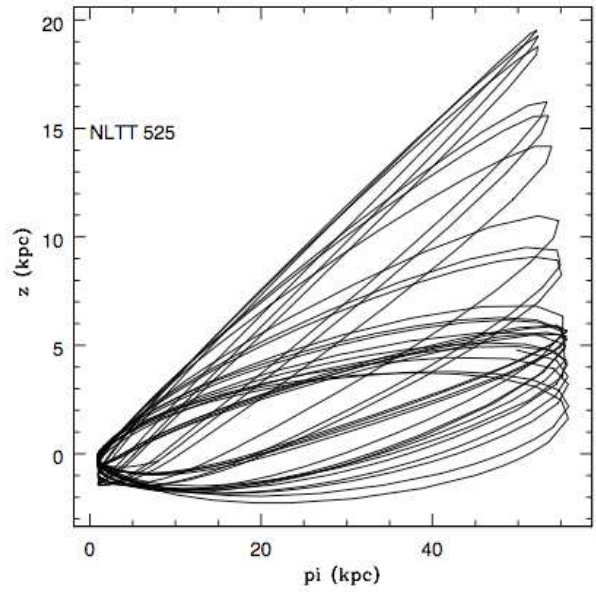}
\figcaption[]{Meridional galactic orbit of the binary NLTT 525/526.  This binary spends 15.7\% of its lifetime within the galactic disk.  }
\end{figure}

\begin{figure}[tbh]
\centering
\includegraphics[width=8cm,height=8cm]{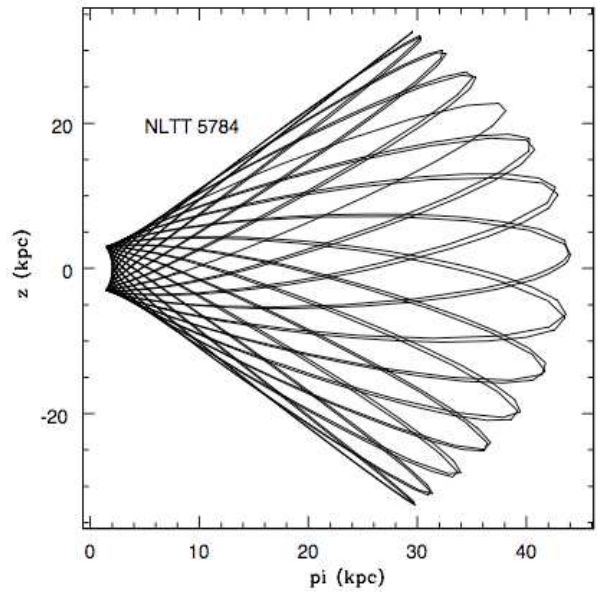}
\figcaption[]{Meridional galactic orbit of the binary NLTT 5781/5784.  This binary spends 1.9\% of its lifetime within the galactic disk.}
\end{figure}

\begin{figure}[tbh]
\centering
\includegraphics[width=8cm,height=8cm]{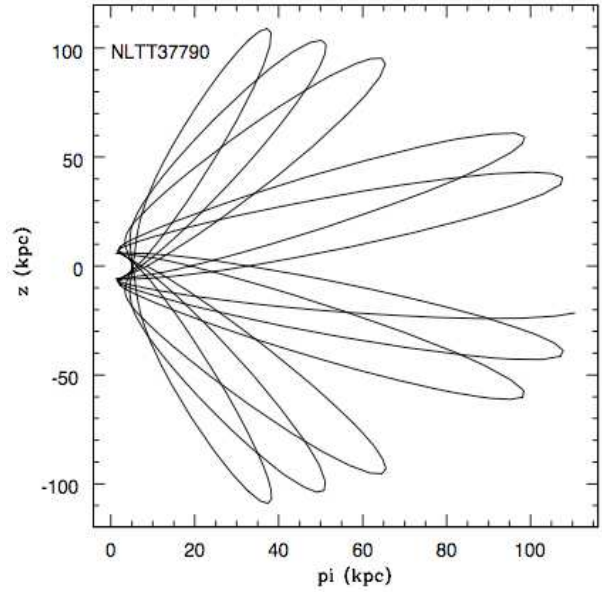}
\figcaption[]{Meridional galactic orbit of the binary NLTT 37790/37787.  This binary spends 0.3\% of its lifetime within the galactic disk.}
\end{figure}

\begin{figure}[tbh]
\centering
\includegraphics[width=8cm,height=8cm]{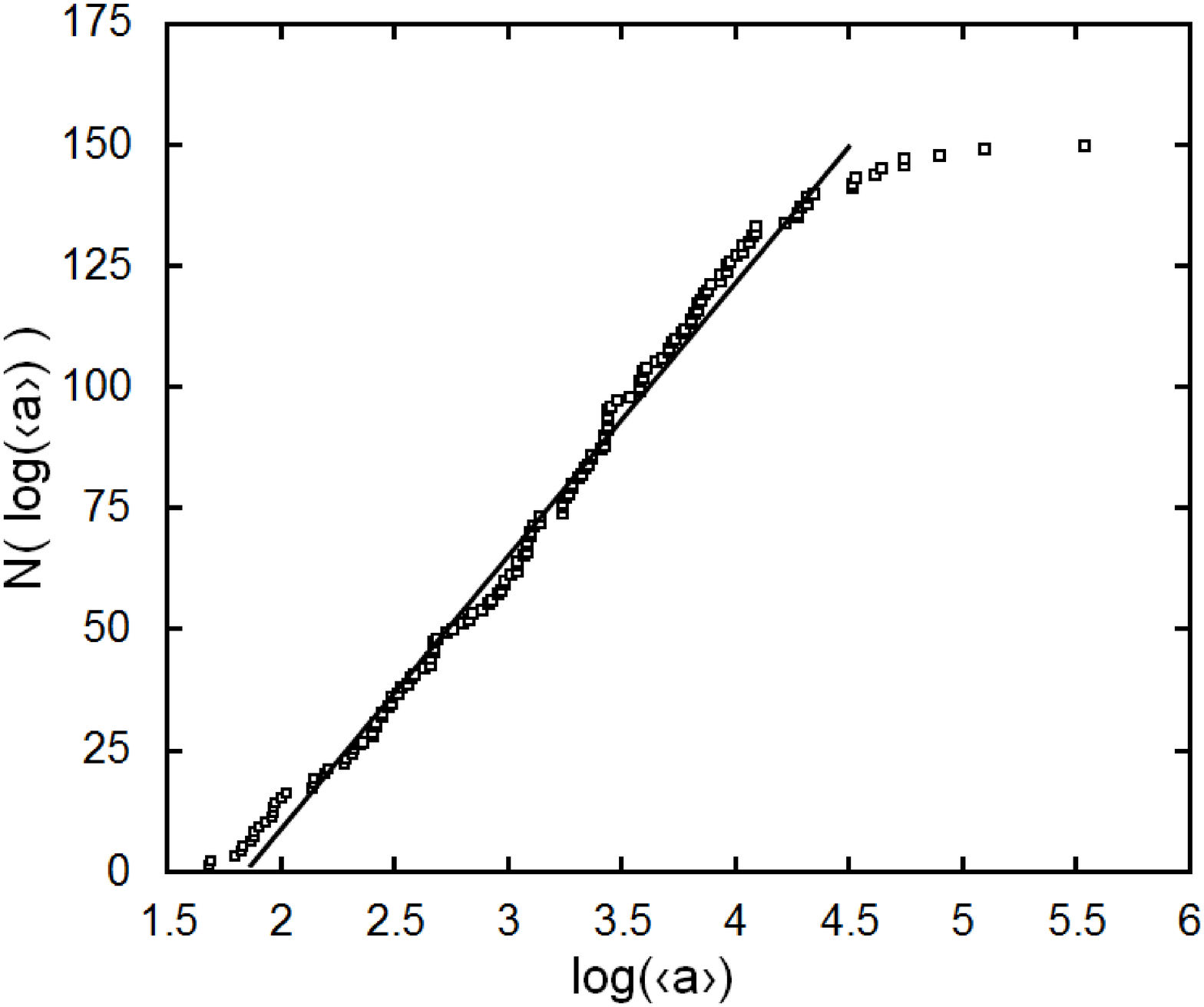}
\figcaption[]{Cumulative distribution of the expected major semiaxes of 150 halo wide binary candidates with computed galactic orbits. Units for $\langle a\rangle$ are AU. The full line is a fit to the Oepik distribution $f(\langle a \rangle) = k/(\langle a \rangle)$}.
\end{figure}

\begin{figure}[tbh]
\centering
\includegraphics[width=8cm,height=8cm]{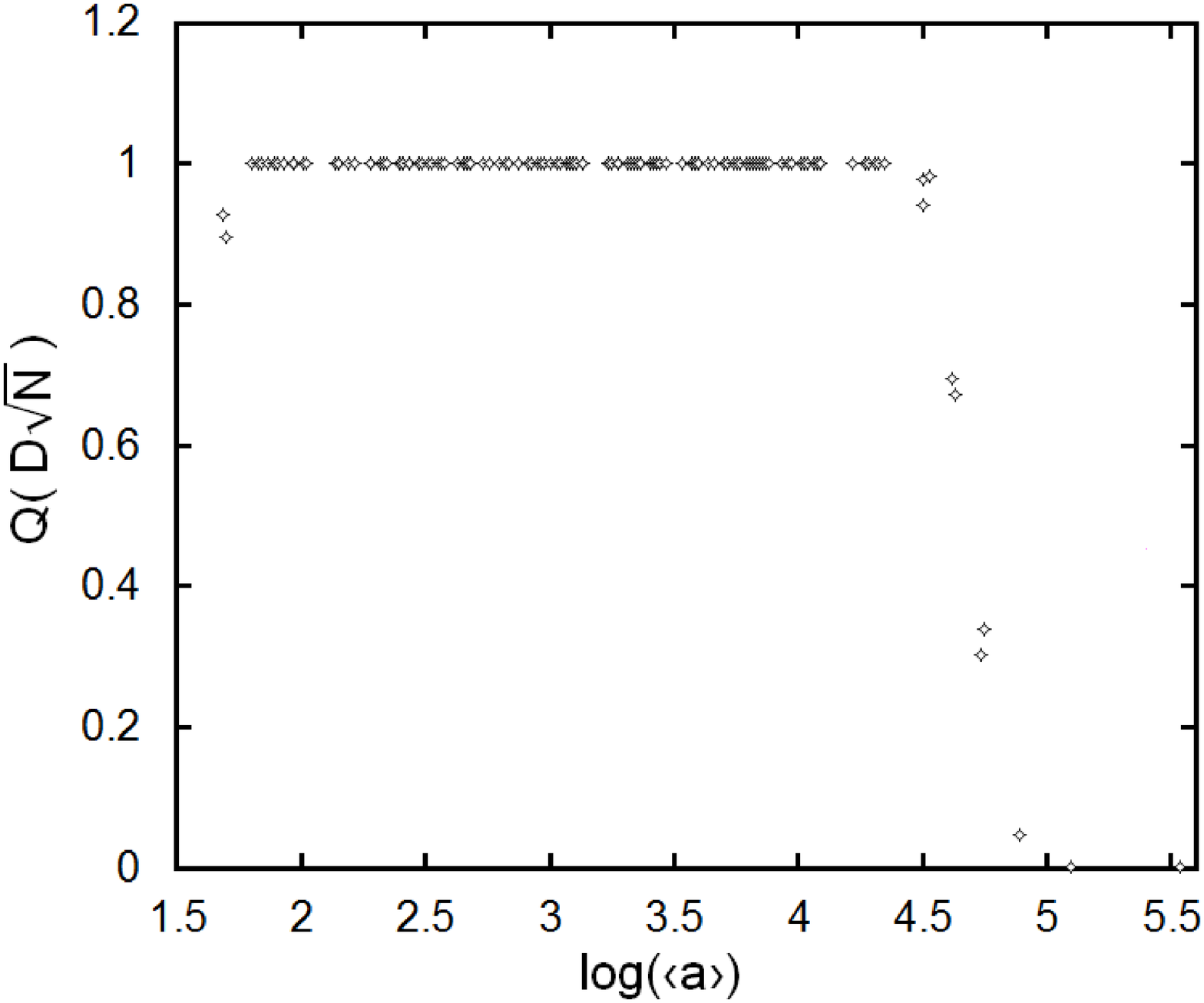}
\figcaption[]{Result of applying the Kolmogorov-Smirnov test to the150 binaries of Figure 14.    The test shows that the fit is excellent  up to an expected major semiaxis of 28,200~AU, and then  abruptly deviates from the Oepik distribution.  We interpret this departure as due to dynamical effects which tend to dissociate the widest binaries.  See text for details.}
\end{figure}

\begin{figure}[tbh]
\centering
\includegraphics[width=8cm,height=8cm]{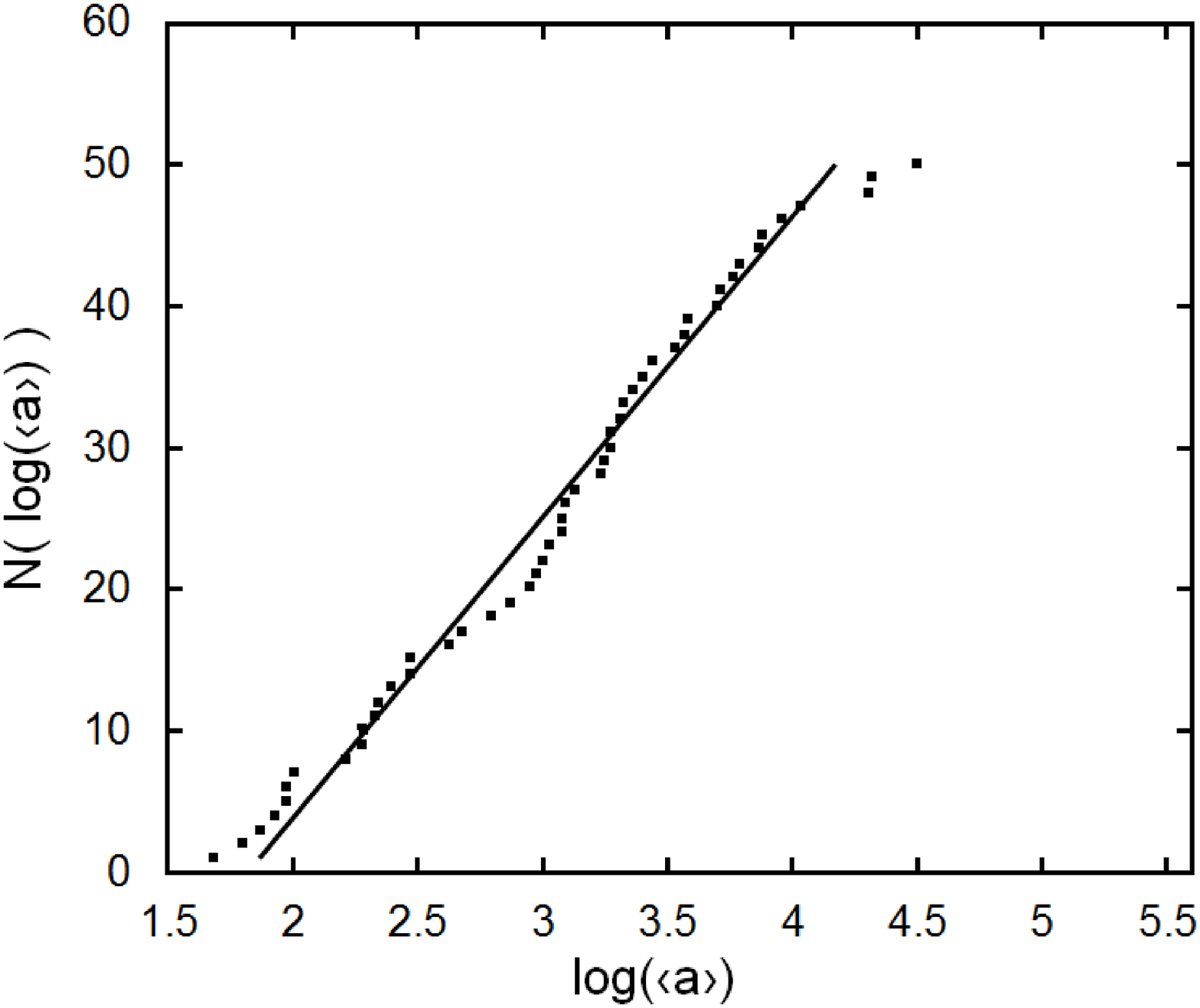}
\figcaption[]{Cumulative distribution of the expected major semiaxes of 50 most disk-like wide binaries  with computed galactic orbits. These binaries spend just about their entire lifetimes within the galactic disk. Units for $\langle a\rangle$ are AU. The full line is a fit to the Oepik distribution $f(\langle a \rangle) = k/(\langle a \rangle)$}.
\end{figure}

\begin{figure}[tbh]
\centering
\includegraphics[width=8cm,height=8cm]{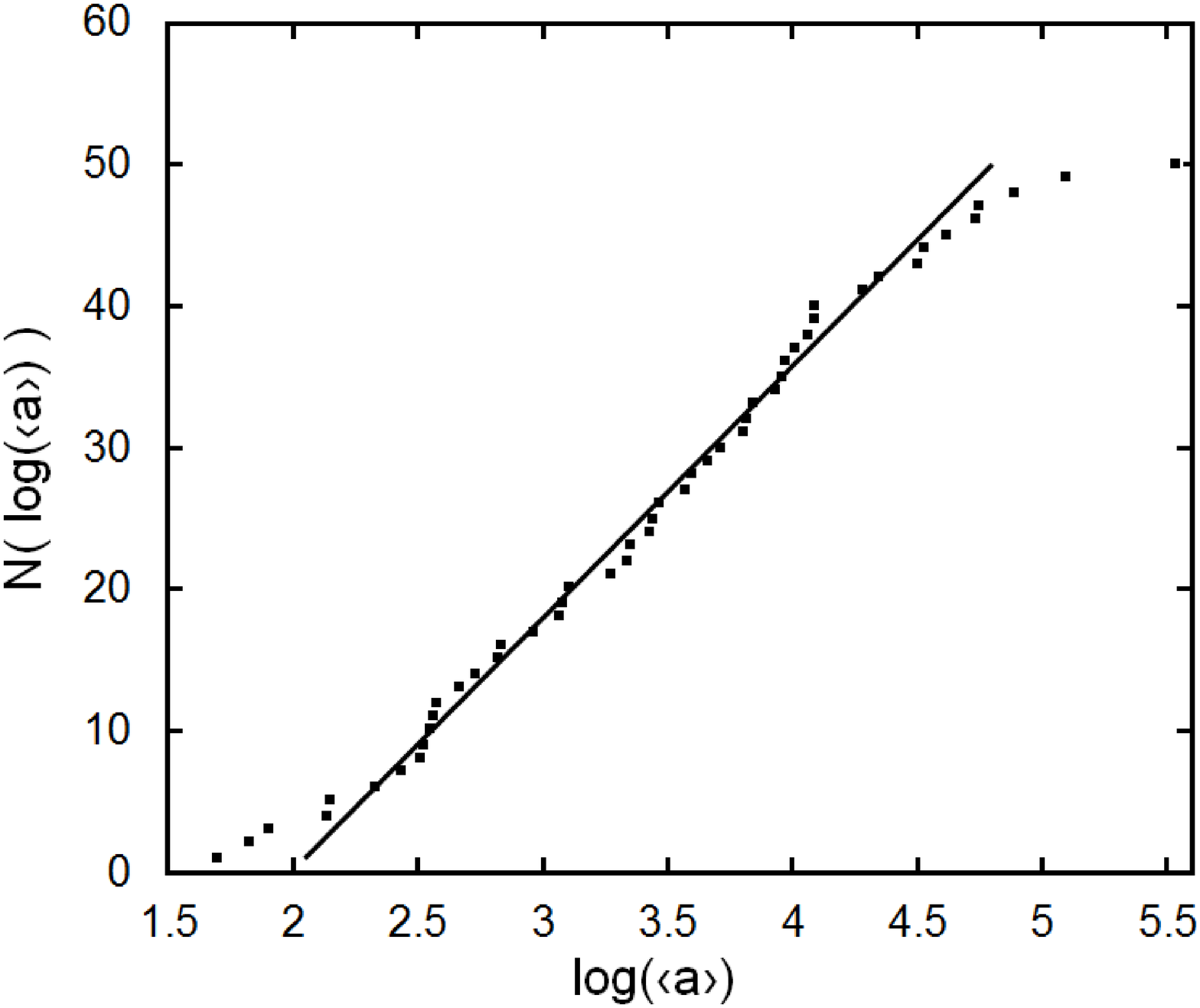}
\figcaption[]{Cumulative distribution of the expected major semiaxes of 50 most halo-like wide binaries  with computed galactic orbits. These binaries spend an average of 18\% of their lifetimes within the galactic disk. Units for $\langle a\rangle$ are AU. The full line is a fit to the Oepik distribution $f(\langle a \rangle) = k/(\langle a \rangle)$.}
\end{figure}

\begin{figure}[tbh]
\centering
\includegraphics[width=8cm,height=8cm]{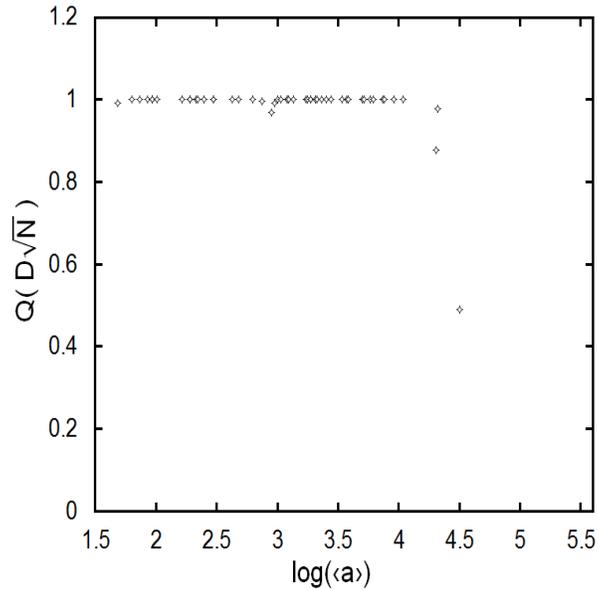}
\figcaption[]{Result of applying the Kolmogorov-Smirnov test to the 50 most disk-like binaries of Figure 16. Units for $\langle a\rangle$ are AU. The test shows that the fit is excellent  up to an expected major semiaxis of 15~000 AU, and then  abruptly deviates from the Oepik distribution.}
\end{figure}

\begin{figure}[tbh]
\centering
\includegraphics[width=8cm,height=8cm]{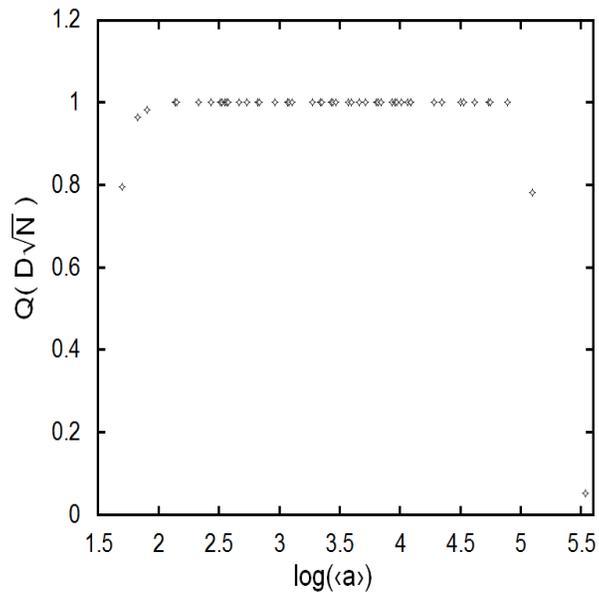}
\figcaption[]{Result of applying the Kolmogorov-Smirnov test to the 50 most halo -like binaries of Figure 17. Units for $\langle a\rangle$ are AU. The test shows that the fit is excellent  up to an expected major semiaxis of 63~000 AU, and then  abruptly deviates from the Oepik distribution.  This result shows that Oepik's distribution holds up to much larger values of the semiaxes as the binaries spend shorter fractions of their lifetimes within the galactic disk.}
\end{figure}

\clearpage

\begin{deluxetable}{lccc cccc cccc cr}

\tabletypesize{\scriptsize} \tablewidth{0pc}
\tablecaption{Improved Catalog of Candidate Halo Wide Binaries}
\tablehead{ 
Primary & Other ID & Secondary & $d$ & $M_V (p)$ & $M_V$(s)& s & $\langle a\rangle$ & $v_p $ & $R_{max}$ & $|z_{max}|$ & $e$ & $t_d/t$ & Cat\\ &&&(pc)& &
&(")&(AU)&(km/s)&(kpc)&(pc)\\
(1)&(2)&(3)&(4)&(5)&(6)&(7)&(8)&(9) &(10)&(11)&(12)&(13)&(14)}
\startdata
NLTT 58746&&NLTT 58747&156&9.4&9.4&4&876&&&&&&CG\\
NLTT 58812&&NLTT 58813&435&6.2&6.2&2&1217&&&&&&CG\\
NLTT 239&&NLTT 240&496&10.0&10.0&10&6944&&&&&&CG\\
NLTT 403&HIP 754&LP 348-62&45&4.5&8.8&30&1902&33.4&8.6&258&0.11&1.000&A\\
NLTT 525&HIP 911&NLTT 526&251&4.8&7.4&8.5&2986&361.3&31.6&3122&0.92&0.157&ACG\\
NLTT 1036&&NLTT 1038&290&10.8&6.0&22.5&9135&213.0&9.8&7363&0.29&0.049&CG\\
NLTT 1412&&NLTT 1416&104&9.5&6.7&51.5&7475&&&&&&CG\\
NLTT 1645&&NLTT 1647&219&7.5&9.9&20.2&6198&&&&&&CG\\
NLTT 1806&HIP 2663&BD -35 170B&36&3.6&6.0&6&304&90.9&9.3&146&0.34&1.000&A\\
NLTT 1945&G 069-004&G 069-004 Ng&53&4.8&5.6&6.16&455&139.5&8.6&921&0.42&0.463&Z\\
NLTT 2056&G172-016&G172-016B&119&5.6&11.4&8.3&1383&123.7&9.5&523&0.70&0.969&Z\\
NLTT 2144&&NLTT 2145&218&6.5&6.5&4&1220&&&&&&CG\\
NLTT 2205&&NLTT 2206&170&6.8&8.2&3&713&&&&&&CG\\
NLTT 2324&&NLTT 2325&198&8.5&8.5&2.5&694&&&&&&CG\\
NLTT 2425&&NLTT 2426&282&11.3&8.7&4&1580&&&&&&CG\\
NLTT 2484&HIP 3550&LP 466-33&63&5.8&13.4&62&5467&69.8&9.7&350&0.22&1.000&A\\
NLTT 2963&&NLTT 2964&397&10.7&11.3&2&1112&&&&&&CG\\
NLTT 2989&&NLTT 2990&107&9.9&9.9&1&149&&&&&&CG\\
NLTT 3116&G 242-074 A&G 242-074 Bn&98&4.8&9.9&19.5&3182&124.5&8.6&1336 &0.52&0.284&Z\\
NLTT 3252&&NLTT 3253&503&9.5&10.7&10.8&7607&&&&&&CG\\
         &HIP 5140&LDS 34&82&4.4&7.8&23&2640&53.5&8.6&528&0.16&0.932&A\\
NLTT 3847&&NLTT 3849&101&8.4&10.9&8&1127&&&&&&CG\\
NLTT 4362&HIP 6130&BD -1 167B&28&5.8&9.7&28&1084&49.7&8.6&112&0.21&1.000&A\\
NLTT 4625&&NLTT 4626&802&6.6&4.4&4&4492&&&&&&CG\\
NLTT 4814&G 002-038&NLTT 4817&153&10.7&5.2&24.4&5139&303.0&12.9&8058&0.98&0.006&ACG\\
NLTT 5690&HIP 7869&LP 88-69&71&4.1&9.1&22&2198&183.7&8.6&4562&0.51&0.083&A\\
NLTT 5781&&NLTT 5784&306&11.6&7.1&52.1&22320&380.4&44.2&32637 &0.91&0.019&CG\\
NLTT 6529&HIP 9094&LP 709-81&34&3.8&9.3&29&1379&80.5&12.4&201&0.26&1.000&A\\
NLTT 7087&HIP 9971&BD 27 335B&57&6.2&6.8&22&1755&86.5&8.6&679&0.32&0.709&A\\
NLTT
7795&G4-16&LP
579-18&134&7.4&12.3&41&7689&147.8&8.6&468&0.67&1.000&A\\
NLTT
7845&HIP 11137&BD 14
387&57&5.1&5.6&34&2727&65.4&12.6&657&0.22&0.730&A\\
NLTT
8447&HIP 12114&BD 6
389B&7&6.5&12.4&170&1715&78.5&11.6&736&0.22&0.580&A\\
NLTT
8643&HIP 12411&LP
941-143&40&5.3&13.4&31&1738&81.1&8.9&311&0.32&1.000&A\\
NLTT
8720&&NLTT 8721&205&7.2&7.2&6&1719&&&&&&CG\\
NLTT 8753&&NLTT
8759&206&9.8&10.9&146.3&42122&&&&&&CG\\
NLTT 8997&HIP 13081&LP
354-414&22&5.8&13.8&20&626&17.7&8.6&68&0.07&1.000&A\\
NLTT
9360&&NLTT 9361&216&10.3&10.1&18.3&5534&&&&&&CG\\ NLTT 9798&HIP
14342&LP 470-9&100&3.6&7.6&1&140&104.5&8.7&2366&0.10&0.121&A\\
NLTT 9961&&NLTT 9962&122&8.7&8.7&4&681&&&&&&CG\\ NLTT 10356&HIP
15126&BD 0
549&79&5.7&11.3&78&8686&155.1&12.7&785&0.48&0.644&A\\ NLTT
10426&HIP 15253&BD 4
519&78&4.8&7.3&2&218&117.8&9.9&76&0.43&1.000&A\\ NLTT 10597&HIP
15371&BD -63
217&12&4.8&5.2&310&5241&74.0&9.5&321&0.24&1.000&A\\ NLTT
10514&HIP 15394&BD 14
550B&57&3.6&10.7&135&10772&72.4&8.8&673&0.23&0.669&A\\ NLTT
10536&HIP 15396&NLTT
10548&214&4.6&9.1&185.7&55410&417.5&9.6&5202&0.12&0.062&ACG\\
NLTT 10634&HIP 15572&BD 8
496&51&4.9&11.5&81&5752&60.5&8.7&444&0.21&1.000&A\\ NLTT
10701&HIP 15683&LP
412-34&73&6.6&13.2&26&2656&92.7&9.3&593&0.32&0.844&A\\ NLTT
10733&HD 20835&LP
155-70&79&4.3&10.5&10&1106&87.2&12.7&817&0.26&0.541&A\\ NLTT
10780&&NLTT 10781&481&8.7&8.7&3&2020&&&&&&CG\\ NLTT 10999&&NLTT
11000&678&8.4&7.6&14&13288&&&&&&CG\\ NLTT 11015&&NLTT
11016&152&10.0&11.1&9&1920&&&&&&CG\\ NLTT 11139&&NLTT
11140&113&6.7&6.5&4&631&&&&&&CG\\ NLTT 11124&HIP 16467&BD 43
744B&54&4.9&9.4&16&1205&70.2&8.9&318&0.27&1.000&A\\ NLTT
11330&L 91-93&L91-94&34&8.4&8.8&13&448&&&&&&R\\ NLTT
11288&&NLTT 11300&176&8.4&5.2&223.5&55082&&&&&&CG\\ NLTT
11424&G6-22&LP
413-23&108&6.1&6.3&3&453&143.6&10.5&891&0.48&0.528&A\\ NLTT
11538&&NLTT 11539&366&4.0&4.0&6&3072&&&&&&CG\\ NLTT 11791&HIP
17666&NLTT
11792&25&6.9&6.3&8&274&151.4&9.6&1603&0.51&0.242&ACG\\ NLTT
12294&&NLTT 12296&189&11.9&8.7&40.5&10691&&&&&&CG\\ NLTT
12415&HIP 18824&LP
944-99&52&3.1&3.7&1.9&139&86.9&8.7&1094&0.25&0.341&A\\ NLTT
12415&HIP
18824&&52&3.1&14.9&64&4678&86.9&8.7&1094&0.25&0.341&A\\ NLTT
12567&HIP
19255&&21&5.6&3.8&720&20739&24.2&9.0&27&0.08&1.000&A\\ NLTT
12567&HIP 19255&BD 37
878B&21&5.6&7.4&3.6&74&24.2&9.0&27&0.08&1.000&A\\ NLTT
12863&HIP 19849&BD -7
781A&5&5.9&10.8&82&578&111.2&12.7&624&0.34&0.815&A\\ &Lp
230-119&Lp 230-125&84&8.5&8.5&31&2606&&&&&&R\\ NLTT 13455&HD
283702&BD 26
727B&80&6.4&8.5&16&1791&92.1&8.6&144&0.41&1.000&A\\ NLTT
13968&&NLTT 13970&148&5.7&12.9&20.3&4209&&&&&&CG\\ NLTT
14005&G85-17&NLTT
13996&103&5.7&12.6&133&19173&101.9&8.7&2000&0.20&0.157&A\\ NLTT
14142&HD 31208&BD 7
754B&34&5.5&5.8&16&761&53.0&8.7&41&0.22&1.000&A\\ NLTT
14407&&NLTT 14408&47&8.5&11.2&11&725&&&&&&CG\\ NLTT 14658&&NLTT
14659&174&5.8&5.8&2.5&609&&&&&&CG\\ NLTT 14867&HIP 24819&BD -3
1061B&17&6.6&6.6&4&94&90.6&9.7&363&0.31&1.000&A\\ NLTT
14943&HIP 25137&LP
638-4&62&5.3&9.1&76&6591&80.6&12.4&1196&0.22&0.307&A\\ NLTT
15008&&NLTT 15009&144&10.4&12.9&12.7&2554&&&&&&CG\\ NLTT
15113&G 99-8&G99-9&114&8.0&9.5&83&9467&&&&&&R\\ NLTT 15262&HIP
26018&BD 19 953&46&5.5&7.3&4&258&56.5&8.7&607&0.16&0.766&A\\
NLTT 15426&HIP 26501&BD -46
1936B&25&5.3&7.9&5.5&194&63.1&10.4&253&0.19&1.000&A\\ NLTT
15512&HIP 26907&BD 2
1041B&31&6.1&11.2&53&2325&88.9&8.5&130&0.40&1.000&A\\ NLTT
15501&&NLTT 15509&210&10.6&11.0&141&41454&328.6&28.2&1481
&0.85&0.374&CG\\ NLTT 15664&&NLTT
15665&45&11.8&11.8&3.5&222&&&&&&CG\\ NLTT 15953&&NLTT
15954&122&10.4&13.3&14.4&2466&&&&&&CG\\ NLTT 15973&HIP
28671&LTT 9457&39&6.4&9.8&7&382&258.6&16.4&1822&0.76&0.220&A\\
NLTT 16062&HIP 28940&BD 34
567&70&6.6&10.4&12&1176&265.5&11.8&6401&0.94&0.103&A\\ NLTT
16184&&NLTT 16185&457&5.9&7.4&4.5&2879&&&&&&CG\\ NLTT
16394&&NLTT 16407&348&4.5&7.6&698.5&340309&481.5&144.0&88786
&0.90&0.003&CG\\ NLTT 16629&&NLTT
16631&55&6.4&10.5&434.1&33416&179.6&8.9&2562 &0.29&0.125&CG\\
NLTT 16747&HIP 31740&LP
160-41&84&5.5&13.7&62&7280&131.8&9.9&784&0.46&0.619&A\\ NLTT
16948&HIP 32423&R
419&25&6.8&11.0&31&1084&70.7&13.7&916&0.24&0.458&A\\ NLTT
17017&HIP 32650&LDS
5686&50&4.4&12.4&27&1893&128.3&12.6&1116&0.37&0.376&A\\ NLTT
17116&G 192-57&LP122-11&238&5.4&8.3&103&24538&&&&&&R\\ NLTT
17382&HIP 34065&BD -43
2907&16&4.5&5.6&21&478&75.7&8.8&156&0.30&1.000&A\\ NLTT
17382&HIP 34065&BD -43
2904&16&4.5&8.1&193&4389&75.7&8.8&156&0.30&1.000&A\\ NLTT
17675&HIP 35599&LP
58-160&85&4.3&11.9&58&6868&117.9&10.0&175&0.42&1.000&A\\ NLTT
17770&HIP 35756&LP
359-227&180&4.1&10.4&34&8565&262.7&11.7&8661&0.96&0.158&A\\
NLTT 17890&HIP 36165&BD -34
3610&40&4.0&4.6&17&944&76.5&9.4&686&0.22&0.635&A\\ NLTT
17907&HIP
36357&&18&2.9&10.6&2.8&69&39.6&10.9&259&0.14&1.000&A\\ NLTT
17907&HIP 36357&BD+32 1562
&18&6.5&2.9&757&18594&39.6&10.9&259&0.14&1.000&A\\ NLTT
18247&&NLTT 18248&110&6.6&6.2&2&307&&&&&&CG\\ NLTT 18346&HIP
37671&NLTT
18347&167&5.1&4.3&11.9&2805&290.3&21.6&17469&0.52&0.028&ACG\\
NLTT 18775&G
90-036&G090-036B&293&5.4&8.8&1.67&685&515.8&429.1&53116
&0.96&0.020&Z\\ NLTT 18798&&NLTT
18799&154&8.9&4.9&12.9&2788&&&&&&CG\\ NLTT 19053&&NLTT
19054&460&4.3&4.3&4&2578&&&&&&CG\\ NLTT 18924&&NLTT
18931&79&10.3&5.5&110.4&12210&155.0&9.0&1958 &0.44&0.185&CG\\
NLTT 19210&G 40-14&BD 30
4982B&235&4.5&10.8&98&32233&413.2&16.4&1444&0.70&0.327&A\\ NLTT
19979&&NLTT 19980&344&7.9&10.2&19.6&9439&228.1&15.0&8120
&0.92&0.135&CG\\ NLTT 20201&&NLTT
20202&248&11.4&8.1&33.4&11581&&&&&&CG\\ NLTT 20699&&NLTT
20700&136&9.7&6.8&5&950&&&&&&CG\\ NLTT 20892&&NLTT
20894&270&9.2&8.0&8&3023&&&&&&CG\\ NLTT 20979&G
09-047&G009-047B&89&2.9&11.6&81.93&10217&162.8&10.1&1617
&0.62&0.233&Z\\ NLTT
21061&G116-009&G116-009B&265&7.2&11.2&10.08&3744&105.7&9.6&5395
&0.89&0.237&Z\\ NLTT 21351&&NLTT
21352&699&3.6&3.6&2.5&2445&&&&&&CG\\ NLTT 21914&&NLTT
21915&420&10.1&9.2&8&4703&&&&&&CG\\ NLTT 21999&HIP
46853&&14&2.5&12.9&5&94&56.2&9.3&220&0.18&1.000&A\\ NLTT
22230&HIP 47225&&34&4.3&9.4&4&191&49.9&8.6&437&0.16&1.000&A\\
NLTT 22301&&NLTT 22302&378&9.3&7.8&13.6&7187&&&&&&CG\\ NLTT
22525&&NLTT 22526&416&6.2&4.5&5&2908&&&&&&CG\\ NLTT 22726&G
43-7&LP
548-41&120&6.3&13.0&15&2519&178.4&13.9&340&0.56&1.000&A\\ NLTT
23500&HIP
49668&AB&38&4.4&8.1&4.7&248&55.5&8.7&620&0.15&0.705&A\\ NLTT
23895&BD+23 2207A&BD+23
2207Bv&23&4.0&9.6&7.72&245&206.9&9.3&150 &0.16&1.000&Z\\ NLTT
23937&&NLTT 23938&289&9.7&9.7&1&405&&&&&&CG\\ NLTT 24638&&NLTT
24639&441&5.9&9.9&6&3700&&&&&&CG\\ NLTT 25233&&NLTT
25234&183&9.8&6.3&9&2300&&&&&&CG\\ NLTT 25403&&NLTT
25404&177&6.8&5.2&25.9&6415&&&&&&CG\\ NLTT 25582&HIP
53165&&55&4.4&6.1&1&77&73.3&9.7&403&0.23&1.000&A\\ NLTT
25623&&NLTT 25624&247&8.4&8.4&3&1038&&&&&&CG\\ NLTT 26709&&NLTT
26710&85&12.0&12.0&1.5&178&&&&&&CG\\ NLTT 27069&G 45-48&BD 6
2436B&92&5.4&8.9&18&2318&103.7&14.5&602&0.33&0.853&A\\ NLTT
27182&&NLTT 27188&128&9.4&7.8&39.1&7030&&&&&&CG\\ NLTT
28236&G176-046&G176-046D&127&7.1&12.0&4.8&853&100.3&8.7&989
&0.63&0.517&Z\\ NLTT 28289&&NLTT
28290&120&10.3&11.1&10&1686&&&&&&CG\\ NLTT 28512&HIP 57443&VB
5&9&5.1&15.7&23&297&58.4&9.4&139&0.20&1.000&A\\ NLTT 29142&HIP
58401&LP 158-35&32&6.4&13.4&23&1027&189.1&8.5&24&0.88&1.000&A\\
NLTT 29424&&NLTT 29425&481&9.3&9.3&3&2020&&&&&&CG\\ NLTT
29556&HIP 58962&BD 19
1185B&135&4.7&7.8&1.8&340&266.9&9.5&5962&0.93&0.232&A\\ NLTT
29553&HIP 58962&NLTT
29556&135&7.8&4.5&11&2267&266.9&9.5&5962&0.93&0.232&ACG\\ NLTT
29665&HIP 59126&LP
494-38&42&6.5&15.0&66&3836&71.5&8.9&206&0.28&1.000&A\\ NLTT
29742&HIP 59233&LP
376-84&56&5.8&10.3&16&1254&91.6&10.5&329&0.29&1.000&A\\ NLTT
29798&&NLTT 29799&240&7.7&7.7&2&672&&&&&&CG\\ NLTT 29860&&NLTT
29862&371&5.0&5.0&7&3635&&&&&&CG\\ NLTT 29957&HIP 59532&LPM
416AB&34&4.8&10.3&2&95&113.3&13.2&743&0.34&0.662&A\\ NLTT
30746&&NLTT 30750&169&10.4&10.8&19.4&4593&&&&&&CG\\ NLTT
30792&&NLTT 30795&370&10.5&7.8&146.7&75889&&&&&&CG\\ NLTT
30837&&NLTT 30838&340&11.4&4.9&15.7&7473&&&&&&CG\\ NLTT 31465&G
59-032&G059-032B&48&5.5&&112.03&7535&211.1&10.6&325
&0.32&1.000&Z\\ NLTT 31648&HIP 62041&LP
377-97&75&6.5&13.7&12&1260&68.6&8.6&375&0.28&1.000&A\\ NLTT
31891&&NLTT 31892&227&7.4&7.4&4&1271&&&&&&CG\\ NLTT 31917&&NLTT
31918&208&8.9&8.9&3&873&&&&&&CG\\ NLTT 32187&HIP 62882&VBS
2&80&3.8&7.8&1.9&213&356.1&22.0&9995&0.94&0.055&A\\ NLTT
32423&HIP 63239&LP
436-66&73&5.5&4.7&58&5934&84.5&11.6&320&0.26&1.000&A\\ NLTT
33175&HIP 64345&LDS
3735&61&4.8&9.1&82&7015&164.4&14.6&1898&0.47&0.217&A\\ NLTT
33209&HIP 64386&LDS
947&70&5.6&7.1&446&43652&117.5&11.0&511&0.37&0.980&A\\ NLTT
33282&&NLTT 33283&188&8.7&9.2&10&2635&&&&&&CG\\ NLTT 33527&HIP
64797&BD 17 2611B&11&6.3&10.1&4&63&48.8&11.1&50&0.17&1.000&A\\
NLTT 34609&G255-038 A&G 255-038
Bq&45&6.6&10.3&14.34&901&183.1&9.3&327 &0.29&1.000&Z\\ NLTT
35210&HIP 67246&LDS
3101&31&3.9&7.5&486&20828&68.0&9.1&195&0.25&1.000&A\\ NLTT
35299&HIP 67408&BD -35
9019B&28&4.3&8.2&12&475&59.8&8.8&33&0.24&1.000&A\\ NLTT
35953&&NLTT 35954&452&10.6&9.8&10.5&6641&&&&&&CG\\ NLTT
35991&&NLTT 36000&52&11.7&9.1&133.4&9792&&&&&&CG\\ NLTT
36418&HIP 69220&BD -44
9130&49&4.9&11.0&4&273&79.8&9.0&795&0.24&0.511&A\\ NLTT
36418&HIP 69220&BD 44
9130&49&4.9&7.0&57&3893&79.8&9.0&795&0.24&0.511&A\\ NLTT
36674&&NLTT 36675&174&9.4&9.4&2&487&&&&&&CG\\ NLTT 37675&&NLTT
37676&162&9.3&9.3&2&454&&&&&&CG\\ NLTT 37787&&NLTT
37790&193&9.9&6.4&200.9&54283&470.0&115.2&109015
&0.92&0.003&CG\\ NLTT 37935&HIP 71505&LP
500-101&99&4.6&5.7&27&3741&140.5&11.3&158&0.46&1.000&A\\ NLTT
37997&HIP 71735&LDS
485&27&5.2&9.0&490&18825&74.1&11.1&922&0.19&0.422&A\\ NLTT
38115&&NLTT 38116&427&3.9&3.9&6&3583&&&&&&CG\\ NLTT 38195&G
66-022&BD +06 2932 B&73&6.1&9.0&3.6&443&267.6&18.1&10301
&0.96&0.137&Z\\ NLTT 39117&&NLTT
39119&375&11.1&5.7&56.7&29736&&&&&&CG\\ NLTT 39442&HIP 74199&LP
424-27&132&6.4&6.6&677&125073&306.7&15.6&7088&0.87&0.075&A\\
NLTT 39456&HIP 74235&NLTT
39457&29&6.7&7.1&300.7&12380&592.2&67.0&6410&0.86&0.088&ACG\\
NLTT 39616&HIP 74549&BD -40
9410&51&4.9&5.0&17&1213&81.6&9.4&216&0.28&1.000&A\\ NLTT
39941&HIP 75069&Ross
1050&45&6.0&9.3&190&11856&58.9&9.5&656&0.14&0.638&A\\ NLTT
40737&&NLTT 40738&540&8.9&8.9&1.6&1209&&&&&&CG\\ NLTT
40901&&NLTT 40902&90&10.0&12.0&4&501&&&&&&CG\\ NLTT 41756&G
16-25&VB 6&256&6.3&8.7&1.5&537&515.9&19.9&6958&0.43&0.060&A\\
NLTT 42554&&NLTT 42555&262&7.7&10.2&5&1832&&&&&&CG\\ NLTT
42846&HIP
80630&&66&5.5&11.0&23&2125&186.8&10.8&313&0.67&1.000&A\\ NLTT
42969&LP745-22&LP745-23&31&9.1&9.6&126&3905&&&&&&R\\ NLTT
43097&G153-067&G
017-027j&48&6.2&10.5&1170.7&79139&162.7&9.4&5692
&0.69&0.074&Z\\ NLTT 43594&&NLTT
43595&158&6.6&8.0&6&1331&&&&&&CG\\ NLTT 44127&HIP 83591&BD -4
4226&11&7.5&9.4&184&2770&74.8&8.9&300&0.28&1.000&A\\ NLTT
44256&&NLTT 44257&139&11.2&11.2&1.5&292&&&&&&CG\\ NLTT
44816&HIP 85378&NLTT
44814&69&4.3&5.5&67&6463&128.9&9.0&1489&0.43&0.253&A\\ NLTT
45000&&NLTT 45001&332&11.4&11.1&9.5&4408&&&&&&CG\\ NLTT
45605&HIP 87533&&66&3.5&7.0&1&93&133.3&9.2&526&0.51&0.955&A\\
NLTT 45995&HIP
88937&&40&4.6&10.1&4.6&260&56.5&8.5&210&0.23&1.000&A\\ NLTT
46029&HIP 89000&&47&2.3&8.3&7.2&473&36.2&8.5&223&0.13&1.000&A\\
NLTT 46029&HIP
89000&&47&2.3&6.6&104&6808&36.2&8.5&223&0.13&1.000&A\\ NLTT
46118&G204-049&G204-049B&86&6.2&10.5&2.7&324&218.3&12.0&1163
&0.50&0.377&Z\\ NLTT 46748&HIP 91360&LP
690-95&37&5.5&11.9&36&1868&76.1&8.5&57&0.35&1.000&A\\ NLTT
46857&HIP
91605&&24&6.8&10.3&9&301&108.2&13.2&484&0.33&1.000&A\\ NLTT
47194&HIP 92960&LP
691-17&75&5.8&10.6&33&3452&145.6&11.3&311&0.47&1.000&A\\ NLTT
47748&&NLTT 47749&254&9.1&8.5&7&2490&&&&&&CG\\ NLTT 47955&HIP
96333&LP 869-10&72&4.7&9.3&40&4031&102.5&8.8&157&0.43&1.000&A\\
NLTT 47955&HIP
96333&&72&4.7&6.7&1.4&141&102.5&8.8&157&0.43&1.000&A\\ NLTT
47984&HIP 96402&&39&4.6&10.8&3&164&64.3&8.7&257&0.25&1.000&A\\
NLTT 48072&&NLTT 48073&44&11.7&11.7&4.5&276&&&&&&CG\\ NLTT
48140&&NLTT 48142&171&4.1&6.6&25.7&6135&&&&&&CG\\ NLTT
48402&HIP 97940&BD 1
4135&47&5.4&5.7&163&10802&57.2&9.1&290&0.19&1.000&A\\ NLTT
48402&HIP
97940&&47&5.4&8.1&76.8&5089&57.2&9.1&290&0.19&1.000&A\\ NLTT
48652&HIP 98767&BD 29
3872B&16&4.7&13.8&178&3959&64.4&8.5&990&0.11&0.346&A\\ NLTT
48686&HIP 99029&&75&4.9&6.6&2&210&98.4&8.9&316&0.39&1.000&A\\
NLTT 48832&HIP 99461&LPM
452&6&6.4&12.5&8&68&130.1&11.3&1139&0.38&0.351&A\\ NLTT
49312&HIP 100970&&37&4.0&7.5&3&157&96.0&9.2&425&0.35&1.000&A\\
NLTT 49486&&NLTT 49487&115&9.8&7.0&4&645&&&&&&CG\\ NLTT
49474&&NLTT 49477&146&5.3&12.4&89.1&18268&&&&&&CG\\ NLTT
49562&G 262-021&G
262-022r&281&6.7&7.2&29.54&12155&370.5&40.0&27828
&0.98&0.031&ZR\\ NLTT 49624&&NLTT
49625&254&7.8&7.8&1.5&533&&&&&&CG\\ NLTT 49746&G 230-047A&G
230-047Bl&78&5.7&10.7&11.63&1159&226.6&10.9&1007
&0.27&0.401&Z\\ NLTT 49819&&NLTT
49821&289&7.1&5.9&25.1&10165&&&&&&CG\\ NLTT 49903&G 230-049A&G
230-049Bm&57&4.7&4.8&1.33&106&234.9&11.7&276 &0.30&1.000&Z\\
NLTT 50559&HIP 104214&BD 38
4344&4&7.5&8.3&24.6&86&95.0&10.1&13&0.32&1.000&A\\ NLTT
50761&HIP
104660&&56&4.3&6.3&1.3&101&142.3&10.9&476&0.48&1.000&A\\ NLTT
50824&&NLTT 50827&375&7.0&4.9&62.7&32910&&&&&&CG\\ NLTT
51353&HIP 106074&Wolf
478&51&7.6&9.1&1.1&80&278.0&8.8&1919&0.80&0.234&A\\ NLTT
51353&HIP 106074&LDS
518&51&7.6&12.1&0.7&50&278.0&8.8&1919&0.80&0.234&A\\ NLTT
51459&G26-8&LP
637-61&72&6.4&9.9&9.5&957&104.9&9.9&52&0.37&1.000&A\\ NLTT
51479&HIP 106335&LP
564-23&49&6.9&8.6&0.7&48&133.1&10.9&97&0.45&1.000&A\\ NLTT
51479&HIP 106335&BD 31
3025&49&6.9&12.5&132&9119&133.1&10.9&97&0.45&1.000&A\\ NLTT
51780&G 93-027&G093-027B&136&6.0&8.7&3.52&672&259.8&14.4&2596
&0.45&0.149&Z\\ NLTT
51964&G188-022&G188-022B&130&4.5&11.4&5.08&922&218.1&12.7&1710
&0.62&0.240&Z\\ NLTT 51992&&NLTT
51993&1138&8.0&7.6&3&4779&&&&&&CG\\ NLTT
52131&G214-001&G214-001B&170&5.9&11.7&5.14&1221&167.1&10.3&1640
&0.64&0.237&Z\\ NLTT 52532&&NLTT
52538&249&7.9&11.9&37.3&12994&&&&&&CG\\ NLTT 52648&&NLTT
52649&954&6.0&6.0&5&6679&&&&&&CG\\ NLTT 52786&&NLTT
52787&167&5.8&5.3&2&468&162.4&9.7&3623 &0.58&0.123&CG\\ NLTT
52864&G
18-24&&119&6.6&11.0&99&16489&201.9&10.4&757&0.76&0.724&A\\ NLTT
53061&&NLTT 53062&39&9.6&9.6&2&110&&&&&&CG\\ NLTT 53254&&NLTT
53255&106&9.2&9.6&53.8&7980&&&&&&CG\\ NLTT 53376&&NLTT
53377&270&9.6&11.8&5&1886&&&&&&CG\\ NLTT 53617&&NLTT
53618&118&10.1&10.1&3&494&&&&&&CG\\ NLTT
54517&LP520-70&LP520-71&107&5.5&8.1&229&24591&&&&&&R\\ NLTT
54708&LP521-3&LP521-4&151&5.2&7.1&38&5741&&&&&&R\\ NLTT
55277&&NLTT 55278&396&5.9&8.6&4&2218&&&&&&CG\\ NLTT 55287&HIP
113231&BD -8
5980B&35&5.3&13.6&42&2047&81.0&9.8&154&0.28&1.000&A\\ NLTT
55326&HIP 113280&BD 16
4838B&32&6.7&8.3&5&225&28.4&8.5&250&0.09&1.000&A\\ NLTT
55719&&NLTT 55720&104&12.7&12.7&3&435&&&&&&CG\\ NLTT 55912&G
028-040&LP 581-80&124&5.8&8.9&15.81&2467&178.9&10.5&230
&0.53&1.000&Z\\ NLTT 55994&&NLTT
55997&234&5.4&10.4&50&16392&&&&&&CG\\ NLTT 56357&HIP 114962&BD
-36 13940B&39&6.2&13.8&15&819&237.9&12.9&575&0.78&0.964&A\\
NLTT 56335&HIP 114980&BD -67
2593&27&6.8&6.6&71&2709&82.5&8.5&380&0.35&1.000&A\\ NLTT
56392&HIP 115012&&55&4.2&5.8&1&77&72.4&8.9&155&0.29&1.000&A\\
NLTT 56856&HIP 115684&LP
346-72&89&4.9&7.2&260&32230&103.4&10.7&334&0.32&1.000&AZ\\ NLTT
56926&G216-045&G216-045B&143&5.5&14.0&31.45&6305&257.1&14.4&94
&0.43&1.000&Z\\ NLTT
57220&G128-077&G128-077B&209&6.8&9.6&9.18&2692&210.8&12.3&1047
&0.59&0.472&Z\\ NLTT 57259&&NLTT
57262&222&12.4&9.6&13.5&4193&&&&&&CG\\ NLTT 57309&HIP
116421&&32&4.3&11.4&8&362&131.2&9.0&1407&0.46&0.269&A\\ NLTT
57558&&NLTT 57559&223&10.7&12.2&15.3&4772&&&&&&CG\\ NLTT
57749&&NLTT 57756&23&10.6&11.3&94.3&3034&&&&&&CG\\ NLTT
57823&&NLTT 57827&433&10.0&8.8&39.5&23942&&&&&&CG\\ NLTT
58235&G273-152&G273-152B&80&5.6&11.3&3.84&430&217.2&10.0&323
&0.20&1.000&Z\\ NLTT 58594&&NLTT
58595&388&8.9&8.9&7&3804&&&&&&CG\\
\enddata

\end{deluxetable}

\end{document}